\documentclass{amsart}

\title
{KP solitons and the Schottky uniformization}

\author{Takashi Ichikawa$^*$ and Yuji Kodama$^{\dagger}$} 
\date{\today}
\address{$^*$ Department of Mathematics, Faculty of Science and Engineering, Saga University, Saga 840-8502, Japan}
\address{
$^\dagger$ Department of Mathematics, The Ohio State University, Columbus, OH 43210, USA}
\email{ichikawn@cc.saga-u.ac.jp}
\email{kodama@math.ohio-state.edu}
\subjclass[2000]{}

\usepackage{url}
\usepackage{color}
\usepackage{rotating}
\newcommand{\rotxc}[1]{\begin{sideways}#1\end{sideways}}
\newcommand{\invert}[1]{\rotxc{\rotxc{#1}}}

\def\Le{\hbox{\invert{$\Gamma$}}}
\countdef\x=23
\countdef\y=24
\countdef\z=25
\countdef\t=26

\def\tbox(#1,#2)#3{
\x=#1 \y=#2 
\multiply\x by 12 
\multiply\y by 12 
\z=\x \t=\y
\advance\z by 12 
\advance\t by 12 
\psline(\x,\y)(\x,\t)(\z,\t)(\z,\y)(\x,\y)
\advance\x by 6
\advance\y by 6 
\rput(\x,\y){{\bf #3}}}

\usepackage{amssymb}
\usepackage{graphicx}
\usepackage{epstopdf}
\usepackage{amscd}
\usepackage{appendix}
\usepackage{graphics}
\usepackage{tikz}
\def\proof{\par{\it Proof}. \ignorespaces}
\def\endproof{{\ \vbox{\hrule\hbox{%
     \vrule height1.3ex\hskip0.8ex\vrule}\hrule }}\par}

\theoremstyle{definition}

\theoremstyle{remark}

\numberwithin{equation}{section}

\let\trueint=\int
\let\truesum=\sum
\def\int{\mathop{\textstyle\trueint}\limits}
\def\sum{\mathop{\textstyle\truesum}\limits}
\let\<=\langle
\let\>=\rangle

\def\RR{{\mathcal R}}

\def\Gr{{\rm Gr}}

\def\SR{\widetilde{\mathcal{R}}_g}

\def\t{\mathbf{t}}
\def\0{\mathbf{0}}
\def\z{\mathbf{z}}

\def\edge{\ar@{-}}
\def\dedge{\ar@{.}}

\usepackage{amsmath, amsthm, amssymb, amsbsy,amsfonts,amscd}
\usepackage[mathscr]{eucal}
\usepackage{epsfig}
\usepackage{young}
\newtheorem{theorem}{Theorem}[section]

\newtheorem{definition}[theorem]{Definition}
\newtheorem{proposition}[theorem]{Proposition}
\newtheorem{lemma}[theorem]{Lemma}

\newtheorem{example}[theorem]{Example}

\newtheorem{remark}[theorem]{Remark}

\usepackage{amsfonts}
\usepackage{xy}
\usepackage{amssymb}
\usepackage{amsmath}

\newcommand{\Z}{\mathbb Z}
\newcommand{\R}{\mathbb R}

\newcommand{\C}{\mathbb C}

\newcommand{\thmrefer}[1]{\renewcommand\thetheorem
  {\protect\ref{#1}}\addtocounter{theorem}{-1}}

\xyoption{all}
\CompileMatrices

\begin{document}

 \begin{abstract}
Real and regular soliton solutions of the KP hierarchy have been classified in terms of
the totally nonnegative (TNN) Grassmannians. These solitons are referred to as KP solitons, and 
they are expressed as singular (tropical) limits of shifted Riemann theta functions.
In this paper, for each element of the TNN Grassmannian, we construct a Schottky group, which 
uniformizes the Riemann surface associated with a real finite-gap solution. Then we 
show that the KP solitons are obtained by degenerating these finite-gap solutions.
\end{abstract}

  \maketitle

\tableofcontents

\section{Introduction}
It is known that  solutions of the KP equation can be constructed
from \emph{any} algebraic curves (Riemann surfaces) \cite{Kr:77}. A solution from a smooth curve is a \emph{quasi-periodic} solution, and some soliton solutions can be  constructed by rational (tropical) limits of the curve with only ordinary double points, i.e. a singular Riemann surface with nodal singularities (see e.g. \cite{Mu:84, SW:85, BB:89}).  In particular, the cases corresponding to the KdV and nonlinear Sch\"odinger equations are well-studied, in which the algebraic curves  are given by the hyperelliptic curves (see e.g.  \cite{BBEIM:94, Mu:84}). Recently, there are several papers dealing with some non-hyperelliptic cases, e.g. so-called $(n,s)$-curves, where the authors construct the Klein $\sigma$-functions over these curves (see e.g. \cite{BLE:97, KMP:18, MK:13, Na:10}).
It seems, however, that almost no result has been reported for the cases with more general algebraic curves.  Because of the difficulty in finding a canonical homological basis for the general algebraic curves, it may be quite complicated to compute
explicitly a rational limit of these curves and the corresponding Riemann theta functions (see \cite{Na:10}).  On the other hand, a large number of \emph{real} and \emph{regular} soliton solutions of the KP hierarchy, referred to as \emph{KP solitons}, has been classified in terms of totally nonnegative (TNN) Grassmannian 
$\text{Gr}(N,M)_{\ge0}$ (see e.g. \cite{KW:13, KW:14, K:17}).  We also mention that 
 recently, there are some progress on the study concerned with the connections between the algebraic-geometric solutions and these soliton solutions \cite{AG:18, Na:18b, Na:24, KX:21}. 
 
 In this paper, we first give a brief review of the KP solitons with combinatorial aspects of the TNN
 Grassmannians. In particular, we describe some details of the so-called \emph{$\Le$-diagram},
 introduced by Postnikov \cite{P:06}, which provides a parametrization of the KP solitons \cite{KW:14}.
 In \cite{K:24}, we identify singular Riemann surfaces for the KP solitons, and introduce the \emph{$M$-theta function} defined on the singular Riemann surface. The $M$-theta function is obtained by singular (rational) limit of the Riemann theta function, 
and it gives the $\tau$-function of the KP soliton.

In order to describe this limiting process 
by the degeneration and deformation theory of Riemann surfaces, 
we use their Schottky uniformization, which provides the explicit description 
of homology cycles, Abelian differentials and period integrals for these uniformized surfaces \cite{Ford, BB:89, Ba:97}.
Based on results of \cite{I:00, I:22}, 
we construct smooth Riemann surfaces uniformized by real Schottky groups  
which degenerate to the above singular Riemann surfaces, 
and then the associated real finite-gap solutions tend to KP solitons. 
Furthermore, 
we show that this degeneration can be interpreted as constructing a smooth deformation
of a Riemann surface with nodal singularities. 
In particular, we show that the $\Le$-diagram in the TNN Grassmannian theory is quite useful 
for the construction. 
More precisely, the $\Le$-diagram can provide the information 
about a canonical homological basis for the smooth Riemann surface. 

The organization of this paper is as follows. 
In Section \ref{sec:RiemannS}, 
following \cite{Mu:84, K:24} 
we recall the definition of the $M$-theta function 
as the limit of the shifted Riemann theta function, 
which express the theta functions of singular Riemann surfaces. 
In Section \ref{sec:KP}, we provide a brief review of the KP solitons and
 the associated $\tau$-function of the KP equation. Then in Section \ref{sec:comb}, we review several combinatorial tools to classify the KP solitons, which include the $\Le$-diagram and 
 pipedream (see e.g. \cite{K:17}). We here also introduce the \emph{$O\Le$-diagram}  (Definition \ref{def:OLe}), which identifies the zero entries of the element $A\in\Gr(N,M)_{\ge0}$.
In Section \ref{sec:Mtheta}, 
we review the result of \cite{K:24} which shows
the $\tau$-functions for the KP solitons as the $M$-theta function. 
We then give in Section \ref{sec:Schottky}  the main result (Theorem \ref{thm:main}) of this paper, which states that KP solitons for TNN Grassmannians can be expressed by 
singular limits of real finite-gap solutions associated with smooth Riemann surfaces 
uniformized by the Schottky groups obtained from 
the combinatorial data, the $O\Le$-diagram. 
Referring to \cite{Ba:97, Bur1:91, Bur2:91, BB:89}, we also give explicit formulas of these finite-gap solutions. 
Finally in Section \ref{sec:Example}, 
we give several explicit examples, and discuss the regularity of the real finite-gap solutions.

\section{The compact Riemann surface and the theta function}\label{sec:RiemannS}
Let $\mathcal{R}_g$ be a smooth compact Riemann surface of genus $g$.
Let $H_1(\mathcal{R}_g,\mathbb{Z})$ be the homology group of $\mathcal{R}_g$, and
a set $\{a_1,\ldots,a_g,b_1,\ldots,b_g\}$ be a canonical basis in $H_1(\mathcal{R}_g,\mathbb{Z})$, that is, we have the intersection products,
\begin{equation*}
a_p\circ a_q=0,\quad b_p\circ b_q=0,\quad a_p\circ b_q=\delta_{p,q}.
\end{equation*}
It is well-known that any compact Riemann surface of genus $g$ is homeomorphic to a sphere with
$g$ handles (see e.g. \cite{FK:91}).
The left panel of the figure below shows a Riemann surface of genus 2.
\begin{figure}[h]
  \centering
  \includegraphics[height=4cm,width=11.5cm]{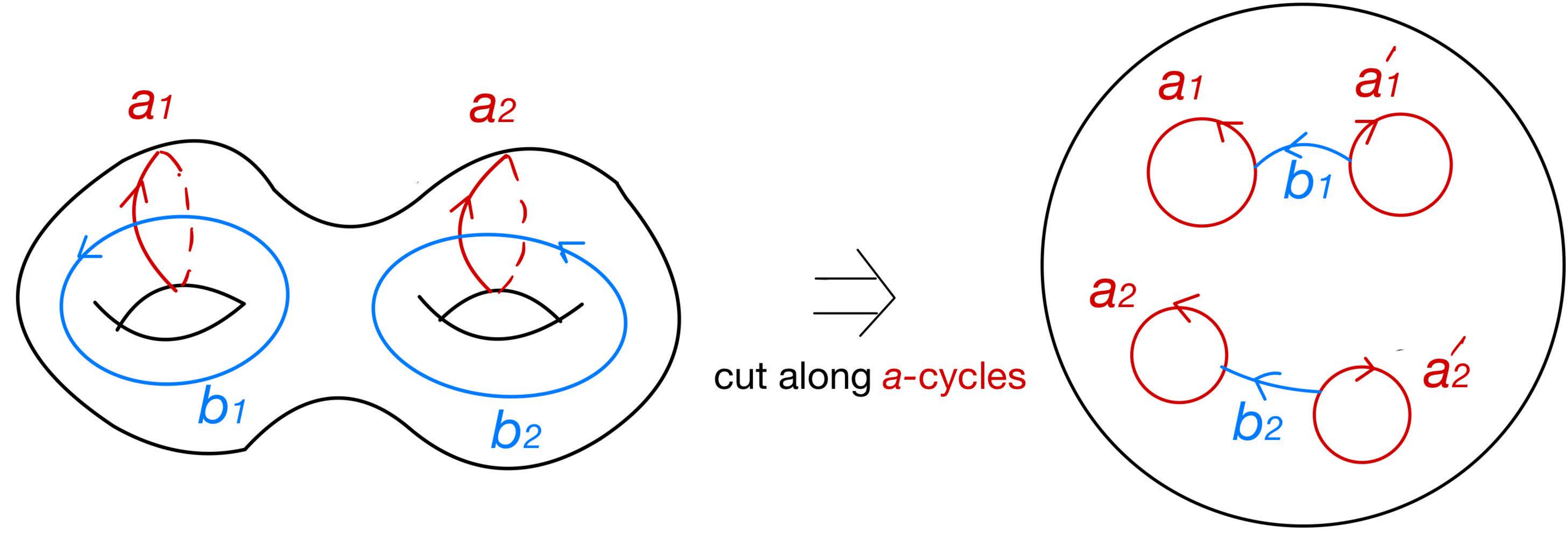}
\end{figure}
Cutting the Riemann surface along the $a$-cycles, we obtain the manifold $\C\mathbb{P}^1$ with
$2g$ holes whose boundaries are marked by $a_p$- and $a'_p$-cycles  as shown in the right panel of the figure. This implies that the Riemann surface can be obtained by identifying each pair of $a_p$- and $a'_p$-cycles. The identification can be expressed by a Schottky group \cite{Ford} as shown in Section \ref{sec:Schottky}, which is the main theme in the present paper.

Given a set of canonical basis of $H_1(\mathcal{R}_g,\mathbb{Z})$, we have the holomorphic
differentials $\{\omega_p:p=1,\ldots,g\}$ normalized by the conditions,
\[
\oint_{a_p}\omega_q=\delta_{p,q},\qquad (1\le p,\,q \le g).
\]
The integrals over the $b$-cycles given by
\begin{equation}\label{eq:Pmatrix}
\Omega_{p,q}:=\oint_{b_p}\omega_q,\qquad (1\le p,\,q\le g)
\end{equation}
define the $g\times g$ period matrix $\Omega=(\Omega_{p,q})$, which is symmetric and $\text{Im}(\Omega)>0$.
Then the Riemann theta function associated with $\mathcal{R}_g$ is defined by
\begin{equation}\label{eq:Riemann}
\vartheta_g({\bf z};\Omega):=\sum_{{\bf m}\in\mathbb{Z}^g}\exp 2\pi i \left(\frac{1}{2}{\bf m}^T\Omega {\bf m}+{\bf m}^T {\bf z}\right),
\end{equation}
for ${\bf z}\in \mathbb{C}^g$, and $\mathbf{m}^T$ is the transpose of the column vector $\mathbf{m}\in\Z^g$. Note that the function $\vartheta({\bf z};\Omega)$ is an entire function on $\C^g$, and $\text{Im}(\Omega)>0$ implies (absolute) convergence at each point on $\C^g$.


\subsection{The $M$-theta function}\label{sec:SRiemann}
In \cite{Mu:84} (Chapter 5, p.3.243), Mumford considered the theta function on singular curve. Let $\widetilde{\mathcal{R}}_g$ be a singular Riemann surface of (arithmetic) genus $g$, and let $S$ be the set of singular points, $S=\{s_1,\ldots,s_g\}\subset \widetilde{\mathcal{R}}_g$.  Assume that the singularities of $\widetilde{\mathcal{R}}_g$ are only ordinary double points $p_1,\ldots,p_g$ and that $\SR$ has normalization
\begin{equation}\label{eq:normal}
\pi:\C\mathbb{P}^1~\longrightarrow~ \SR \quad\text{with}\quad \pi^{-1}(s_p)=\{\alpha_p,\beta_p\}
\end{equation}
That is, $\SR$ is just $\C\mathbb{P}^1$ with $g$ pairs of points $\{\alpha_p,\beta_p\}$ identified. 
Figure below shows the case with $g=2$.
\begin{figure}[h]
  \centering
  \includegraphics[height=4cm,width=11.5cm]{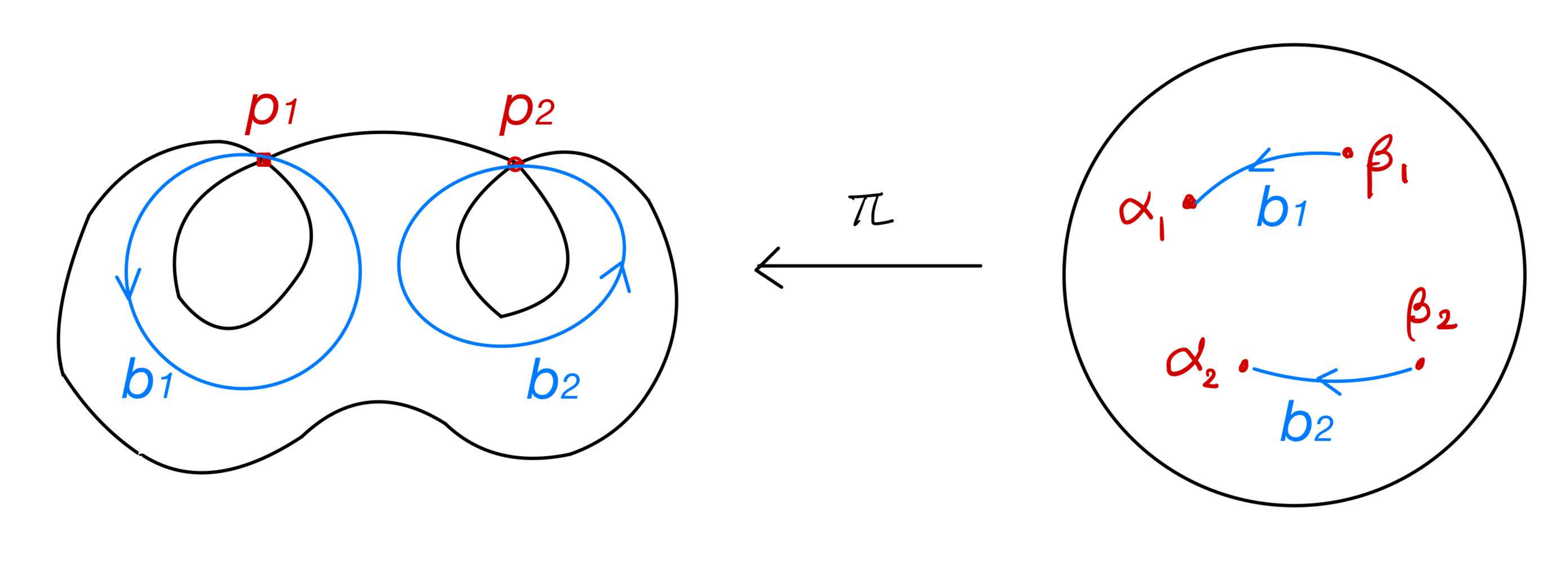}
\end{figure}
The singular Riemann surface $\SR$ is obtained by pinching all $a$-cycles as shown in the figure.

By pinching $a$-cycles, the holomorphic differentials $\{\omega_p:p=1,\ldots,g\}$ take the limits \cite{BB:89, Ka:11,I:22} (see also Section \ref{sec:deformation}, where we show this limit),
\begin{equation}\label{eq:Lomega}
\omega_p\quad\longrightarrow\quad \widetilde{\omega}_p=\frac{dz}{2\pi i}\left(\frac{1}{z-\alpha_p}-\frac{1}{z-\beta_p}\right).
\end{equation}
Then the limits of the period matrix in \eqref{eq:Pmatrix} are calculated as
\begin{equation}\label{eq:LPmatrix}
\Omega_{p,q}\quad\longrightarrow\quad \widetilde{\Omega}_{p,q}:=\int_{\beta_p}^{\alpha_p}\widetilde{\omega}_q=\frac{1}{2\pi i}\ln\,C_{p,q} \qquad \text{mod}(\Z),
\end{equation}
where $C_{p,q}=\exp(2\pi i\widetilde{\Omega}_{p.q})$ is given by the cross-ratio $[\alpha_p,\beta_p;\alpha_q,\beta_q]$, 
\begin{equation}\label{eq:CR}
C_{p,q}=[\alpha_p,\beta_p;\alpha_q,\beta_q]:=\frac{(\alpha_p-\alpha_q)(\beta_p-\beta_q)}{(\alpha_p-\beta_q)(\beta_p-\alpha_q)}.
\end{equation}
Note in particular that the diagonal parts of the period matrix $\Omega$ has the limits
\begin{equation}\label{eq:limits}
\text{Im}~\Omega_{p,p}~ \longrightarrow ~+\infty \quad \text{for}\quad 1\le p\le g.
\end{equation}
Then the limit of the $\vartheta$-function \eqref{eq:Riemann} is just $1$, which corresponds to the choice ${\bf m}^T=(0,\ldots,0)$. To obtain a nontrivial example, we consider the (half) shifts
\[
z_p ~\longrightarrow ~ z_p-\frac{1}{2}\Omega_{p,p},\quad\text{for}\quad 1\le p\le g,
\]
which then gives the Riemann theta function with shifted variable ${\bf z}\in\mathbb{C}^g$,
\begin{equation}\label{eq:RiemannS}
\vartheta_g({\bf z};\Omega)=\sum_{{\bf m}\in\mathbb{Z}^g}\exp 2\pi i\left(\frac{1}{2}\sum_{p=1}^gm_p(m_p-1)\Omega_{p,p}+\sum_{p<q}m_pm_q\Omega_{p,q}+\sum_{p=1}^gm_pz_p\right).
\end{equation}
Then the limit $\Omega_{p,p}\to+i\infty$ for all $p=1,\ldots, g$ leads to
\begin{align}\label{eq:Theta}
\vartheta_g(\z;\Omega)~\longrightarrow~ &~\tilde\vartheta_g(\z;\widetilde\Omega):=\sum_{{\bf m}\in\{0,1\}^g}\exp 2\pi i\left(\sum_{p<q}m_pm_q\widetilde\Omega_{p,q}+\sum_{p=1}^gm_pz_p\right)\\
=&1+\sum_{p=1}^g e^{2\pi i z_p}+\sum_{p<q}C_{p,q}e^{2\pi i (z_p+ z_q)}+\ldots+\left(\prod_{p<q}C_{p,q}\right)e^{2\pi i\sum_{p=1}^g z_p},\nonumber
\end{align}
where $\widetilde{\Omega}$ denotes the limit of $\Omega$ as in \eqref{eq:CR}.
Note that the {infinite} sum of exponential terms in the $\vartheta$-function \eqref{eq:RiemannS} becomes a \emph{finite} sum of $2^g$ exponential terms with $m_p\in\{0,1\}$, if all $C_{p,q}\ne 0$ for $p<q$.
In \cite{K:24}, the function $\tilde\vartheta_g$ is referred to as the \emph{$M$-theta function}.
\begin{remark}\label{rem:Cjk}
When all the pairs $\{(\alpha_p,\beta_p):1\le p\le g\}$ are real and $\alpha_p<\beta_p$ (w.l.o.g.), one should note that the cross ratio $C_{p,q}$ in \eqref{eq:CR} takes the signs depending on the orders of the pairs, i.e.
\begin{itemize}
\item[\rm{(i)}] if $\alpha_p<\beta_p<\alpha_q<\beta_q$ or $\alpha_p<\alpha_q<\beta_q<\beta_p$, then $C_{p,q}>0$,
\item[\rm{(ii)}] if $\alpha_p<\alpha_q<\beta_p<\beta_q$, then $C_{p,q}<0$, and
\item[\rm{(iii)}] if $\alpha_p=\alpha_q$ or/and $\beta_p=\beta_q$, then $C_{p,q}=0$.
\end{itemize}
The case {\rm (ii)} will be important when we discuss the regularity of the soliton solutions (see also {\rm\cite{HKL:24}}). Also note that the case {\rm (iii)} implies that the off-diagonal element $\widetilde{\Omega}_{p,q}$ also takes $+i\infty$, in addition to the diagonal elements in the singular limit \eqref{eq:limits}.
\end{remark}

\begin{remark}\label{rem:partialP}
One can also consider a \emph{partial} pinch with only some $a$-cycles. In this case, we expect that the $\vartheta$-function
associated with the singular Riemann surface may be expressed by a sum of $\vartheta$-functions of
genus \emph{less} that $g$ with phase shifts \cite{K:24, N:20}. In the case with pinching all but except one $a$-cycle, the theta function may be expressed in terms of the Weierstrass functions (see e.g. \cite{Ka:23, LZ:22, YN:13}). We will discuss the details in a future communication \cite{IK:25}.
\end{remark}


\section{The KP equation}\label{sec:KP}
In this section, we give a brief summary of the KP solitons for the purpose of the present paper (see e.g. \cite{K:17} for the details). The KP equation is a nonlinear partial differential equation for a single function $u=u(x,y,t)$ in the form,
\begin{equation}\label{eq:KP}
\partial_x(-4\partial_t u+6u\partial_xu+\partial_x^3u)+3\partial_y^2u=0,
\end{equation}
where $\partial_z^k:=\frac{\partial^k}{\partial z^k}$ for $z=x,y,t$.
The solution of the KP equation is given in the following form,
\begin{equation}\label{eq:u}
u(x,y,t)=2\partial_x^2\ln \tau(x,y,t),
\end{equation}
where $\tau(x,y,t)$ is called the $\tau$-function of the KP equation.

\subsection{KP solitons}
The soliton solutions are constructed as follows:
Let $\{f_i(x,y,t):1\le i\le N\}$ be a set of linearly independent functions $f_i(x,y,t)$  satisfying the following system of linear equations,
\begin{equation}\label{eq:f}
\partial_yf_i=\partial_x^2 f_i,\quad \text{and}\quad  \partial_t f_i=\partial_x^3 f_i\qquad\text{for}\quad  i=1,\ldots,N.
\end{equation}
The Wronskian $\text{Wr}(f_1,\ldots,f_N)$ with respect to the $x$-variable gives a $\tau$-function, that is,
the function $u(x,y,t)$ in \eqref{eq:u} is a solution of the KP equation,
\begin{equation}\label{eq:Wr}
\tau(x,y,t)=\text{Wr}(f_1,f_2,\ldots,f_N).
\end{equation}
(See, e.g. \cite{K:17} for the details.)

As a fundamental set of the solutions of \eqref{eq:f}, we take the exponential functions
$E_j(x,y,t)$ for $j=1,\ldots, M$ $(M>N)$,
\begin{equation}\label{eq:E}
E_j(x,y,t)=e^{\xi_j(x,y,t)}\quad\text{with}\quad \xi_j(x,y,t):=\kappa_jx+\kappa_j^2y+\kappa_j^3t.
\end{equation}
where $\kappa_j$'s are arbitrary constants. In this paper, we consider the real and regular soliton solutions, for which assume $\kappa_j$'s be all real with the ordering
\begin{equation}\label{eq:orderK}
\kappa_1~<~\kappa_2~<~\cdots~<~\kappa_M.
\end{equation}
For the soliton solutions, we consider $f_i(x,y,t)$ as a linear combination of the exponential solutions,
\begin{equation}\label{eq:Ef}
f_i(x,y,t)=\sum_{j=1}^MA_{i,j}E_j(x,y,t)\qquad\text{for}\qquad i=1,\ldots,N.
\end{equation}
where $A:=(A_{i,j})$ is an $N\times M$ constant real matrix of full rank, $\text{rank}(A)=N$.
Then the $\tau$-function \eqref{eq:Wr} is expressed by
\begin{equation}\label{eq:tauE}
\tau(x,y,t)=|AE(x,y,t)^T|,
\end{equation}
where $E(x,y,t)^T$ is the transpose of the $N\times M$ matrix $E(x,y,t)$ defined by
\begin{equation}\label{eq:Et}
E(x,y,t)=\begin{pmatrix}
E_1 & E_2 &\cdots & E_M\\
\kappa_1E_1&\kappa_2E_2&\cdots &\kappa_ME_M\\
\vdots &\vdots &\ddots &\vdots\\
\kappa_1^{N-1}E_1 &\kappa_2^{N-1}E_2&\cdots &\kappa_M^{N-1}E_M
\end{pmatrix}.
\end{equation}
Note here that the set of exponential functions $\{E_1,\ldots,E_M\}$ gives a basis of $M$-dimensional space
of the null space of the operator $\prod_{i=1}^M(\partial_x-\kappa_i)$.  Then the set of functions $\{f_1,\ldots,f_N\}$ represents an $N$-dimensional subspace
of $M$-dimensional space spanned by the exponential functions. This leads naturally to the structure
of a finite real Grassmannian $\text{Gr}(N,M)$, the set of $N$-dimensional subspaces in $\R^M$.
Then the $N\times M$ matrix $A$ of full rank can be identified as a point of $\text{Gr}(N,M)$, and throughout the paper we assume
$A$ to be in the reduced row echelon form (RREF). 
\begin{definition}
An $N\times M$ matrix $A$ in RREF, i.e. $A\in\Gr(N,M)$, is irreducible, if
\begin{itemize}
\item[\rm{(a)}] in each row, there is at least one nonzero element besides the pivot, and
\item[\rm{(b)}]  there is no zero column.
\end{itemize}
This implies that the first pivot is at $(1,i_1)=(1,1)$ entry of $A$, and the last pivot at $(N,i_N)$ with $N\le i_N<M$.
\end{definition}

The $\tau$-function in \eqref{eq:tauE} can be expressed as the following formula using the Binet-Cauchy lemma (see e.g. \cite{K:17}),
\begin{equation}\label{eq:tauPl}
\tau(x,y,t)=\sum_{I\in\binom{[M]}{N}}\Delta_{I}(A)E_{I}(x,y,t),
\end{equation}
where $I=\{i_1<i_2<\cdots<i_N\}$ is an $N$ element subset of $[M]:=\{1,2,\ldots,M\}$, $\Delta_{I}(A)$ is the $N\times N$ minor with the column vectors indexed by $I=\{i_1,\ldots,i_N\}$,
and $E_{I}(x,y,t)$ is the $N\times N$ determinant of the same set of the columns in \eqref{eq:Et}, which is given by
\begin{equation}\label{eq:EI}
E_{I}=\prod_{k<l}(\kappa_{i_l}-\kappa_{i_k})\,E_{i_1}\cdots E_{i_N}=\prod_{k<l}(\kappa_{i_l}-\kappa_{i_k})
\exp\left(\xi_{i_1}+\cdots+\xi_{i_N}\right).
\end{equation}
The minor $\Delta_{I}(A)$'s are the Pl\"ucker coordinates of $\wedge^N\R^M$, and the $\tau$-function represents a
point of $\text{Gr}(N,M)$ in the sense of the Pl\"ucker embedding, $\text{Gr}(N,M)\hookrightarrow \mathbb{P}(\wedge^N\R^M): A\mapsto \{\Delta_I(A):I\in\binom{[M]}{N}\}$. 
It is then obvious that if all the minors of $A$ are nonnegative, the $\tau$-function \eqref{eq:tauPl}
is sign-definite, i.e. the solution $u$ in \eqref{eq:u} is \emph{regular}. The $\text{Gr}(M,N)$
consisting of these elements is called the totally nonnegative (TNN) Grassmannian, denoted by
$\text{Gr}(N,M)_{\ge 0}$. Then the following theorem for the \emph{necessary} condition of the regularity was proven in \cite{KW:13}.
\begin{theorem}\label{thm:regularity}
The soliton solution generated by the $\tau$-function \eqref{eq:tauPl} is regular, \emph{if and only if}
the matrix $A$ is in $\rm{Gr}(N,M)_{\ge0}$.
\end{theorem}
This implies that the KP solitons can be classified by the TNN Grassmannians. That is, given an element $A\in \Gr(N,M)_{\ge0}$, we have a unique KP soliton. The construction of the KP solitons and their asymptotic properties are provided by some combinatorial tools \cite{KW:14}, which we explain below.

\section{Combinatorics for the TNN Grassmannians}\label{sec:comb}
We here provide a brief summary of combinatorial description of the TNN Grassmannian $\text{Gr}(N,M)_{\ge 0}$ (see also \cite{K:17} for the details). Each element $A\in\text{Gr}(N,M)$ is expressed as an $N\times M$ matrix in
the reduced row echelon form. Let $\{i_1,\ldots,i_N\}$ be the pivot set of the matrix $A$. Then 
the Young diagram corresponding to the pivot set is obtained as follows.
Consider a lattice path starting from the top right corner and ending at the bottom left corner with the label $\{1,\ldots,M\}$, so that the pivot indices
appear at the vertical paths as shown in the diagram below. Then each box of the Young diagram can be labeled by the pair of indices $(i,j)$, where $i$ is the pivot index and $j$ the non-pivot index.
\bigskip

\setlength{\unitlength}{0.58mm}
\begin{center}
  \begin{picture}(120,60)
\put(5,55){\line(1,0){142}}
\put(5,45){\line(1,0){110}}
  \put(5,35){\line(1,0){90}}
  \put(5,25){\line(1,0){90}}
  \put(5,15){\line(1,0){70}}
  \put(5,5){\line(1,0){70}}
  \put(5,5){\line(0,1){50}}
 \put(31,5){\line(0,1){50}}
  \put(18,5){\line(0,1){50}}
  \put(75,5){\line(0,1){10}}
  \put(62,5){\line(0,1){10}}
  \put(95,25){\line(0,1){10}}
  \put(82,25){\line(0,1){10}}
\put(115,45){\line(0,1){10}}
\put(102,45){\line(0,1){10}}
  \put(98,29){${i_{n}}$}
  \put(-35,29){$M-N+n$}
  \put(-35,49){$M-N+1$}
  \put(-13,38){$\vdots$}
  \put(-13,18){$\vdots$}
  \put(65,38){$\vdots$}
  \put(10,38){$\vdots$}
  \put(24,38){$\vdots$}
  \put(100,38){${}$}
  \put(60,49){$\cdots$}
  \put(60,58){$\cdots$}
  \put(55,29){$\cdots$}
  \put(47,18){$\vdots$}
  \put(10,18){$\vdots$}
  \put(24,18){$\vdots$}
  \put(80,18){${}$}
 \put(-15,9){$M$}
  \put(45,9){$\cdots$}
  \put(78,9){${i_N}$}
  \put(0,58){$M-N$}
 \put(105,58){${i_1}{\qquad\cdots\qquad 1}$}
  \put(118,49){${i_1}{\quad\cdots\quad 1}$}
 
    \put(7,-3){$ M$}
    \put(18,-3){${M-1\quad\cdots}$}
    \put(60,-3){${i_N+1}$}
 \end{picture} 
\end{center}

\medskip
\noindent
We recall that the partitions $\lambda$ are in bijection with $N$-element
subset $I\subset [M]$, i.e. we have  $\lambda_1\ge\lambda_2\ge\cdots\ge\lambda_N$ with
\[
 \lambda_k=M-N-(i_k-k)\qquad\text{for}\quad k=1,\ldots,N.
\]
The irreducible element $A\in\text{Gr}(N,M)_{\ge 0}$ defines the \emph{irreducible} Young diagram,
which has $\lambda_1=M-N$ and $\lambda_N\ge 1$.


\subsection{The Le-diagram}\label{sec:Le}
In \cite{P:06}, Postnikov introduced the $\Le$-diagram (called \emph{Le}-diagram), which gives a parametrization of 
the element $A\in\text{Gr}(N,M)_{\ge0}$ (more precisely, the $\Le$-diagram uniquely parametrizes the \emph{positroid} cell in Postnikov's decomposition of $\Gr(N,M)_{\ge0}$, to which $A$ belongs \cite{KW:14}).
\begin{definition}\label{def:Le}
A $\Le$-diagram is a decorated Young diagram with \raisebox{0.12cm}{\hskip0.18cm\circle{5}\hskip-0.15cm} in some boxes, which satisfies 
 the property (called {\it{$\Le$-property}}): If there is \raisebox{0.12cm}{\hskip0.18cm\circle{5}\hskip-0.15cm}, then all the boxes either to its left or above it are all \raisebox{0.12cm}{\hskip0.18cm\circle{5}\hskip-0.15cm}.
That is, there is no such \raisebox{0.12cm}{\hskip0.15cm\circle{5}\hskip-0.15cm}, which has an empty box to its left and an empty box above it.  Each \young[1][7][\raisebox{-0.07cm}{\hskip0.18cm\circle{5}\hskip-0.18cm}] corresponds to a vanishing minor $\Delta_I(A)=0$ for some index $I\in\binom{[M]}{N}$ (see Theorem 5.6 in \cite{KW:13}).

We also say that a $\Le$-diagram is irreducible,
if each column and row has at least one empty box (i.e no zero column or/and no zero row). See the left
diagram in Example \ref{ex:LeP59} below.
\end{definition}
Then Postnikov in \cite{P:06} proved the following theorem.
\begin{theorem}\label{thm:Le-S}
There is a bijection between the set of irreducible $\Le$-diagram and the set of derangements of the symmetric group $S_M$.
\end{theorem}
Here the derangement associated to the $\Le$-diagram can be found by constructing a \emph{pipedream}
on the diagram as follows (see \cite{K:17} for the details):  Starting from a $\Le$-diagram, we replace an empty box with a box containing elbow-pipes
connected by a bridge and replace a box with \raisebox{0.12cm}{\hskip0.15cm\circle{5}\hskip-0.15cm} by a box containing crossing pipes as shown below.
\begin{figure}[h]
\centering
\includegraphics[width=8cm]{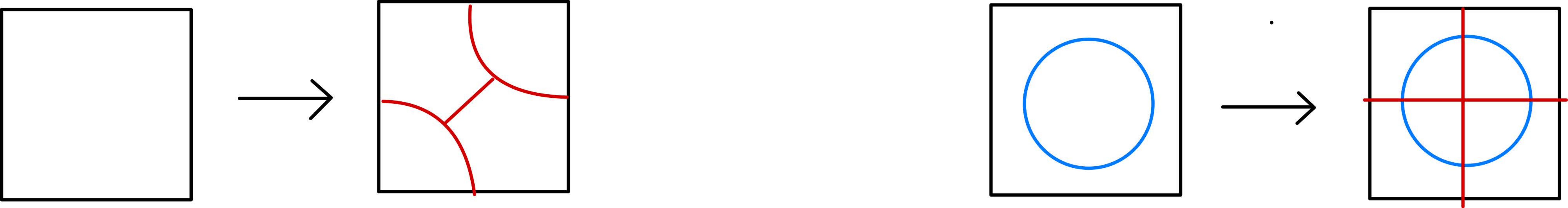}
\end{figure}
Then we label the southeast (input) boundary of the $\Le$-diagram from $1$ to $M$ starting from the top
corner to the bottom corner of the boundary. We place a pipe with the index of the input edge from
the southeast (output) boundary to the northwest boundary,
and then label each northwest edge according to the index of the pipe.
Then the derangement $\sigma$ with a pair $(i,j)$ in
$\sigma(i)=j$ can be found on the opposite sides of the boundary. 
\begin{example}\label{ex:LeP59}
Below shows a $\Le$-diagram and the corresponding pipedream. Then we have the derangement  $\sigma=(8,6,2,5,4,7,9,1,3)$ in one-line notation.
\begin{figure}[h]
  \centering
  \includegraphics[width=8cm]{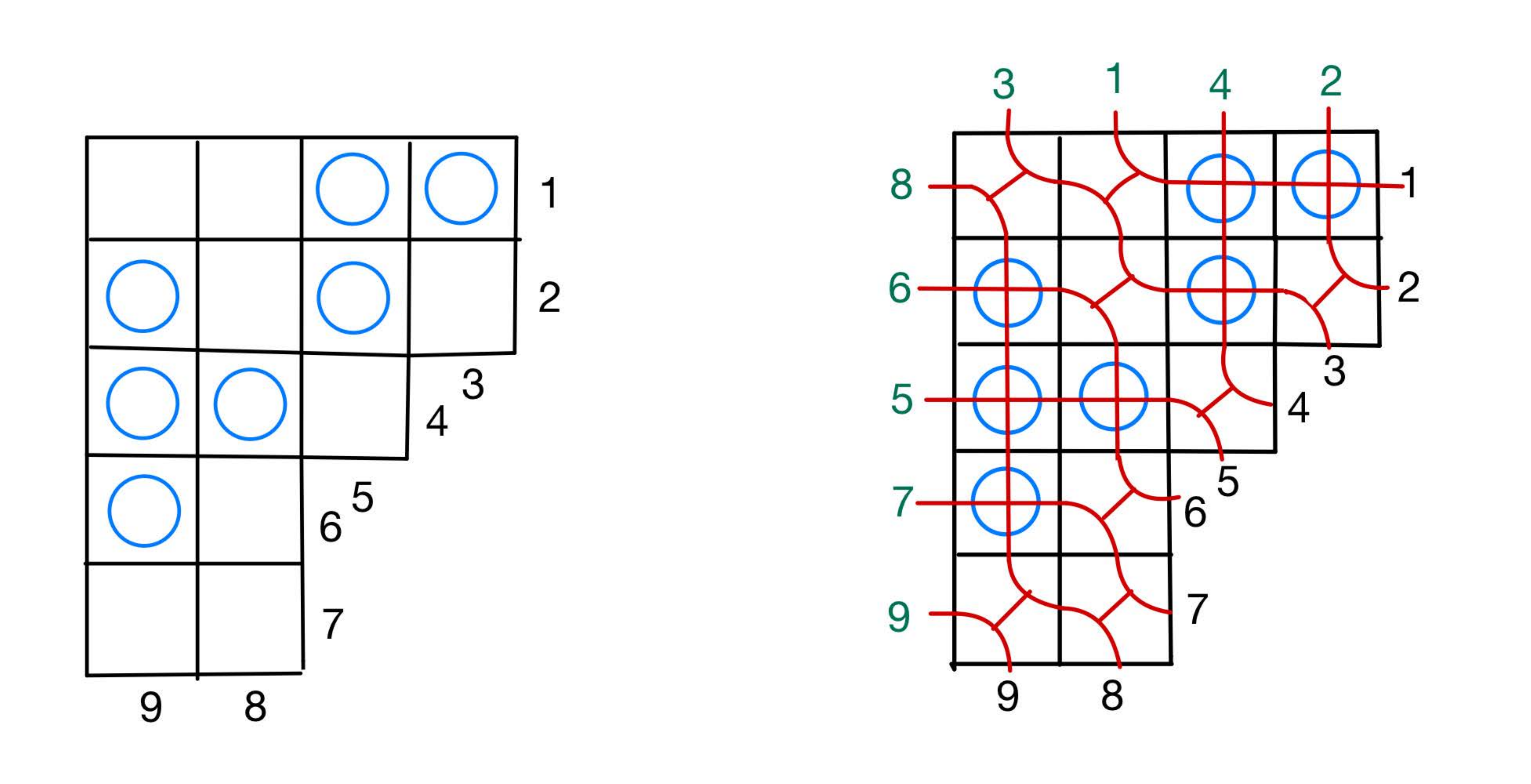}
\end{figure}
\end{example}

 From the $\Le$-diagram, one can find the matrix $A\in\Gr(N,M)_{\ge 0}$ (see e.g. \cite{K:17}). In particular,
we have the following proposition about the zero elements in $A$.
\begin{proposition}\label{prop:Zeros}
Given an irreducible $\Le$-diagram, the zero entries of $A\in\text{Gr}(N,M)_{\ge0}$ can be determined as follows. Consider the box at $(i_k,j)$ with \raisebox{0.12cm}{\hskip0.15cm\circle{5}\hskip-0.15cm}
whose south-east corner is a point on the boundary of the diagram.
Then we have two cases as shown in the figure below.
\begin{itemize}
\item[{\rm (a)}]
The $k$-th row, say $A_{k,\bullet}$, of the matrix $A$ has the structure,
\[
A_{k,\bullet}=(\ldots, 0, 1,\ldots, *, 0, 0,\ldots, 0),
\]
that is, the pivot ``1" is at the $(k,i_k)$ entry of $A$ and the nonzero element marked by ``$*$'' is at the $(k,j-1)$ entry. The entries $A_{k,l}$ for $j\le l\le M$ are all zero.
\item[{\rm (b)}]
The $j$-th column, say $A_{\bullet,j}$, of the matrix $A$ has the structure,
\[
(A_{\bullet,j})^T=(0, 0,\ldots,0,*,\ldots),
\]
that is, the entries $A_{l,j}=0$ for $1\le l\le k$ are all zero, and the nonzero element ``$*$'' is at the $(i_{k+1},j)$ entry with $i_{k+1}=i_k+1$.
\end{itemize}
\begin{figure}[h]
  \centering
  \includegraphics[width=8cm]{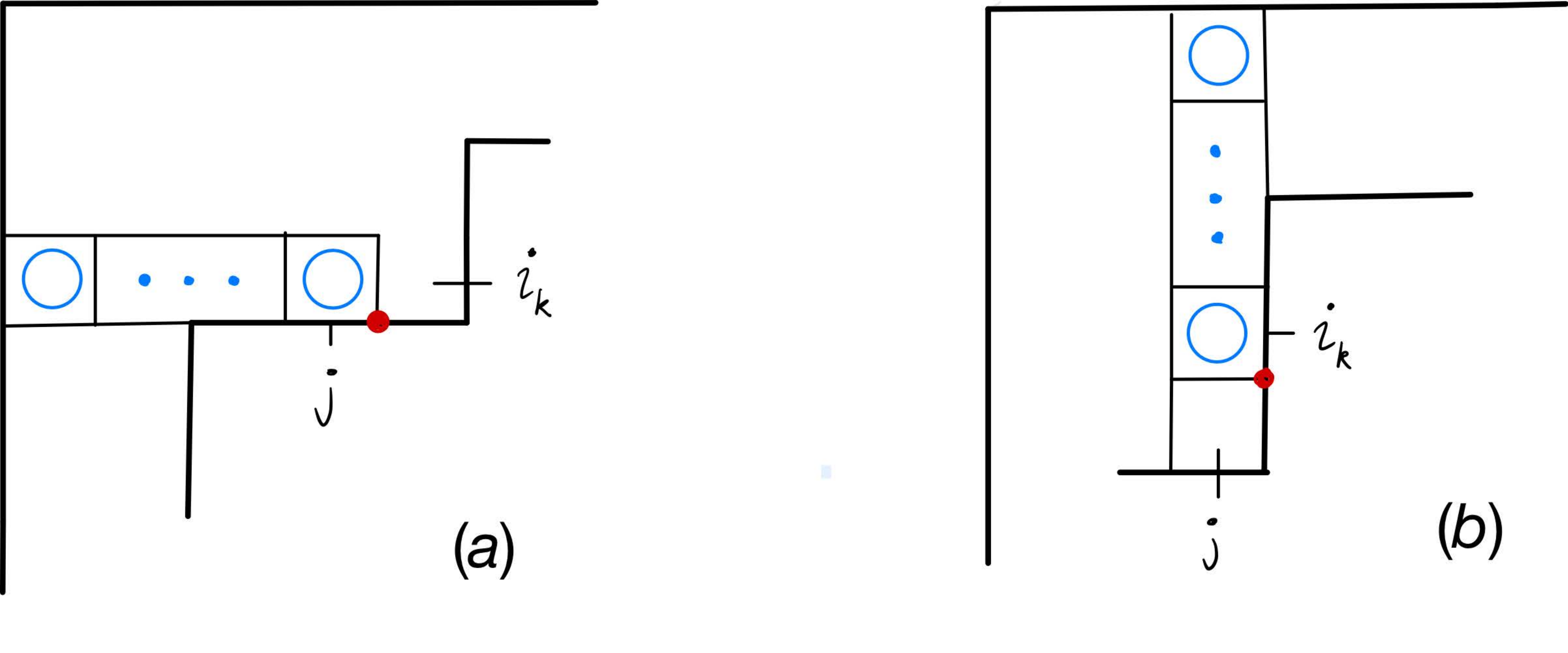}
\end{figure}
\end{proposition}
\proof
Using Theorem 5.6 and Corollary 5.8  in \cite{KW:13} about the vanishing minors (the generalized chamber ansatz), one can show that
\begin{itemize}
\item[(a)] the minor $\Delta_{i_1,\ldots,i_{k-1},j,i_{k+1,\ldots,i_N}}(A)=0$, and 
\item[(b)] the minor $\Delta_{i_1,\ldots,i_{k-1},i_{k+1},\ldots,i_l,j,i_{l+1},\ldots,i_N}(A)=0$,
\end{itemize}
which imply the equations in the proposition, that is, (a) shows $A_{k,l}=0$ for $l\ge j$, and
(b) shows $A_{l,j}=0$ for $l\le k$.
 Note that  there is a case $j>i_{k+1}$ in (a). This can can be also proven in the same way.
\endproof

Each \young[1][7][\raisebox{-0.07cm}{\hskip0.18cm\circle{5}\hskip-0.18cm}] in the $\Le$-diagram gives a vanishing minor of the matrix $A$. Proposition \ref{prop:Zeros} provides the information of the zero entries
of $A$, which are given by the vanishing minors in (a) and (b) in the proof.

\begin{example}\label{ex:P1}
Consider the example \ref{ex:LeP59}. The middle diagram in the figure below
shows the nonzero entries other than pivots $A_{k,i_k}=1$ in the matrix $A$, e.g. $A_{2,8}\ne 0$ and $A_{3,5}\ne0$. Each empty box gives zero entry of $A$, e.g. $A_{1,3}=A_{1,5}=A_{3,8}=0$.
Each star in the middle diagram implies that there is a path $[i,j]$ through the pipedream from the pivot index $i$ at the east boundary to the non-pivot index $j$ at the south boundary of the $\Le$-diagram.
\begin{figure}[h]
\centering
\includegraphics[width=10.5cm]{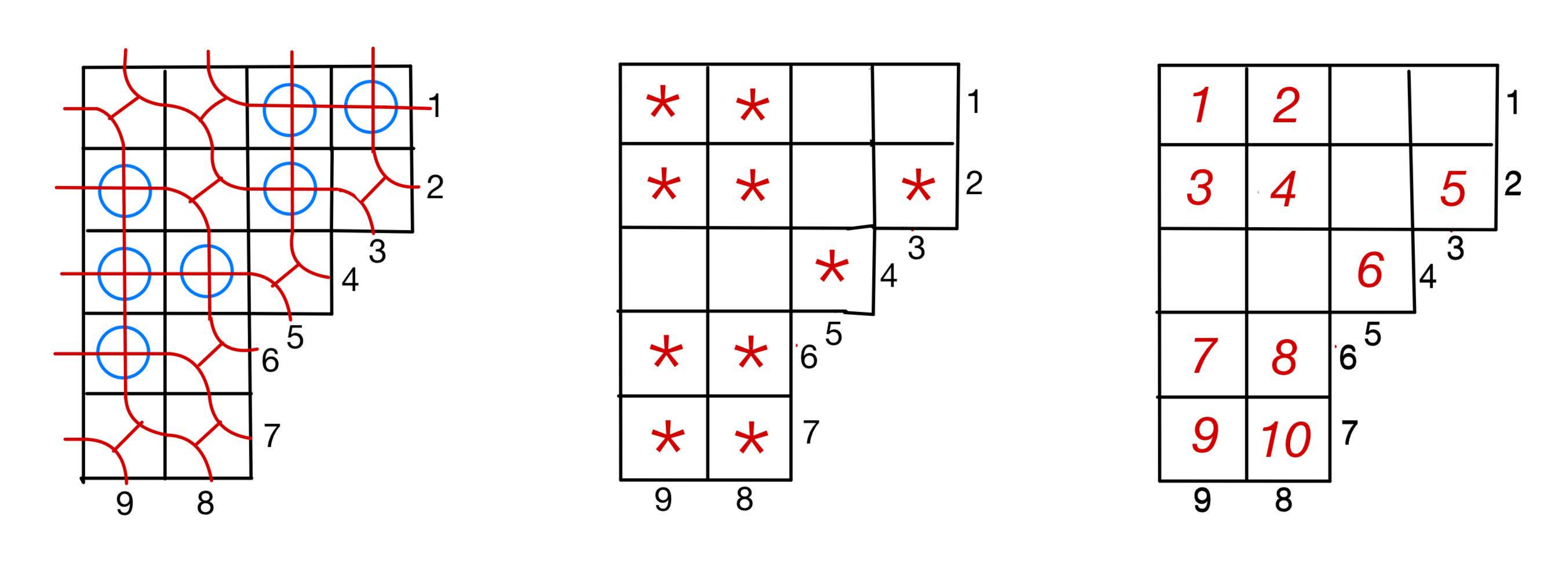}
\end{figure}

Then the matrix $A$ can be expressed in the form,
\[
A=\begin{pmatrix}
1 & 0 & {\bf 0} & 0 & {\bf 0} & 0 & 0 & * & * \\
  & 1  & *  & 0 & {\bf 0} & 0 & 0 & * & *\\
  &    & & 1 & * & 0 & 0 & {\bf 0} & {\bf 0} \\
  & & & & & 1 & 0 & * & * \\
  & & & & & & 1 & * & * 
  \end{pmatrix}
  \]
In the matrix, the bold face ${\bf 0}$'s are corresponding to the boxes with \raisebox{0.12cm}{\hskip0.15cm\circle{5}\hskip-0.15cm} described in Proposition \ref{prop:Zeros}, i.e.
they are at $(1,3), (1,5), (2,5), (4,8)$ and $(4,9)$ in the $\Le$-diagram. The right diagram will be called
the $O\Le$-diagram (see Definition \ref{def:OLe} below).
\end{example}

\subsection{The matroid $\mathcal{M}(A)$ and the set $P_1(A)$}

For $A\in\text{Gr}(N,M)_{\ge 0}$, we define the matroid,
\begin{equation}\label{eq:matroid}
\mathcal{M}(A)=\left\{I\in\binom{[M]}{N}:\Delta_I(A)\right\}.
\end{equation}
Let $I_0$ be the lexicographically minimum element of $\mathcal{M}(A)$. Then we have the decomposition,
\begin{equation}\label{eq:decompM}
\mathcal{M}(A)=\bigcup_{n=0}^N \mathcal{M}_n(A),
\end{equation}
where
\[
\mathcal{M}_n(A):=\left\{J\in\mathcal{M}(A):|J\cap I_0|=N-n\right\}.
\]
For Example \ref{ex:P1}, we have $I_0=\{1,2,4,6,7\}$.
Note that $\mathcal{M}_0(A)=\{I_0\}$.  
We also define $P_1(A)$ as the set of pairs $[i,j]$,
\begin{equation}\label{eq:P1}
P_1(A):=\left\{[i,j]:i\in I_0\setminus J, j\in J\setminus I_0~\text{for}~J\in\mathcal{M}_1(A)\right\}
\end{equation}
This implies that $P_1(A)$ is identified as the set of nonzero entries in $A$ besides the pivots,
that is, $[i_k,j_l]\in P_1(A)$ represents
\begin{itemize}
\item[(a)] $i_k\in I_0\setminus J$ is the $k$-th pivot of $A$, i.e. $A_{k,i_k}=1$,
\item[(b)]  $j_l\in J\setminus I_0$ is the nonzero element $A_{k,j_l}$ in the $k$-th row.
\end{itemize}

We then define an order in $P_1(A)$ and the \emph{$O\Le$-diagram} showing this order in the
$\Le$-diagram.
\begin{definition}\label{def:OLe}
Let $\ell:\,P_1(A)\,\to\, [g]$ with $g:=|P_1(A)|$ be a bijection satisfying the order,
\begin{itemize}
\item[(1)] $\ell([i,k])<\ell([i,l])$, if $k>l$,
\item[(2)] $\ell([i,\bullet])<\ell([j,\bullet])$, if $i<j$.
\end{itemize}
That is, the elements of $P_1(A)$ can be uniquely numbered from 1 to $g=|P_1(A)|$, i.e.
\begin{equation}\label{eq:order}
1~\le ~p=\ell([i,j])~\le ~g,\quad \text{for}~~ [i,j]\in P_1(A).
\end{equation}
The $O\Le$-diagram is defined as the diagram obtained by assigning these numbers to the boxes of the $\Le$-diagram (remove all \raisebox{0.12cm}{\hskip0.18cm\circle{5}\hskip-0.15cm}\,'s). See the right panel in  Example \ref{ex:P1}. Note that the blank boxes in the $O\Le$-diagram are these \young[1][7][\raisebox{-0.07cm}{\hskip0.18cm\circle{5}\hskip-0.18cm}] described in Proposition \ref{prop:Zeros}.
\end{definition}

Following \cite{K:24}, we can identify the singular points $S=\{s_1,\ldots,s_g\}$ where the inverse
image of the normalization \eqref{eq:normal} is given by
\begin{equation}\label{eq:NOP1}
\pi^{-1}(s_p)=\{\alpha_p=\kappa_i, \beta_p=\kappa_j\} \quad \text{for}\quad p=\ell([i,j]) ~\text{and}~1\le p\le g.
\end{equation}
As will be shown in the next section, the number $g$ gives the genus of the Riemann surface associated with the KP soliton.

From the $O\Le$-diagram, we can also show the following proposition on the sign of the coefficient $C_{p,q}$.
\begin{proposition}\label{prop:Cpq}
In the $O\Le$-diagram, consider a rectangular section whose corner boxes are marked $a,b,c$ and $d$ 
with $a<b<c<d$ as shown in the figure below. We also assign a pair of parameters $(\kappa_i,\kappa_l)$ to each box 
according to the boundary indices of the $\Le$-diagram.
Then we have that
\begin{figure}[h]
\hskip1cm
\includegraphics[width=5.5cm]{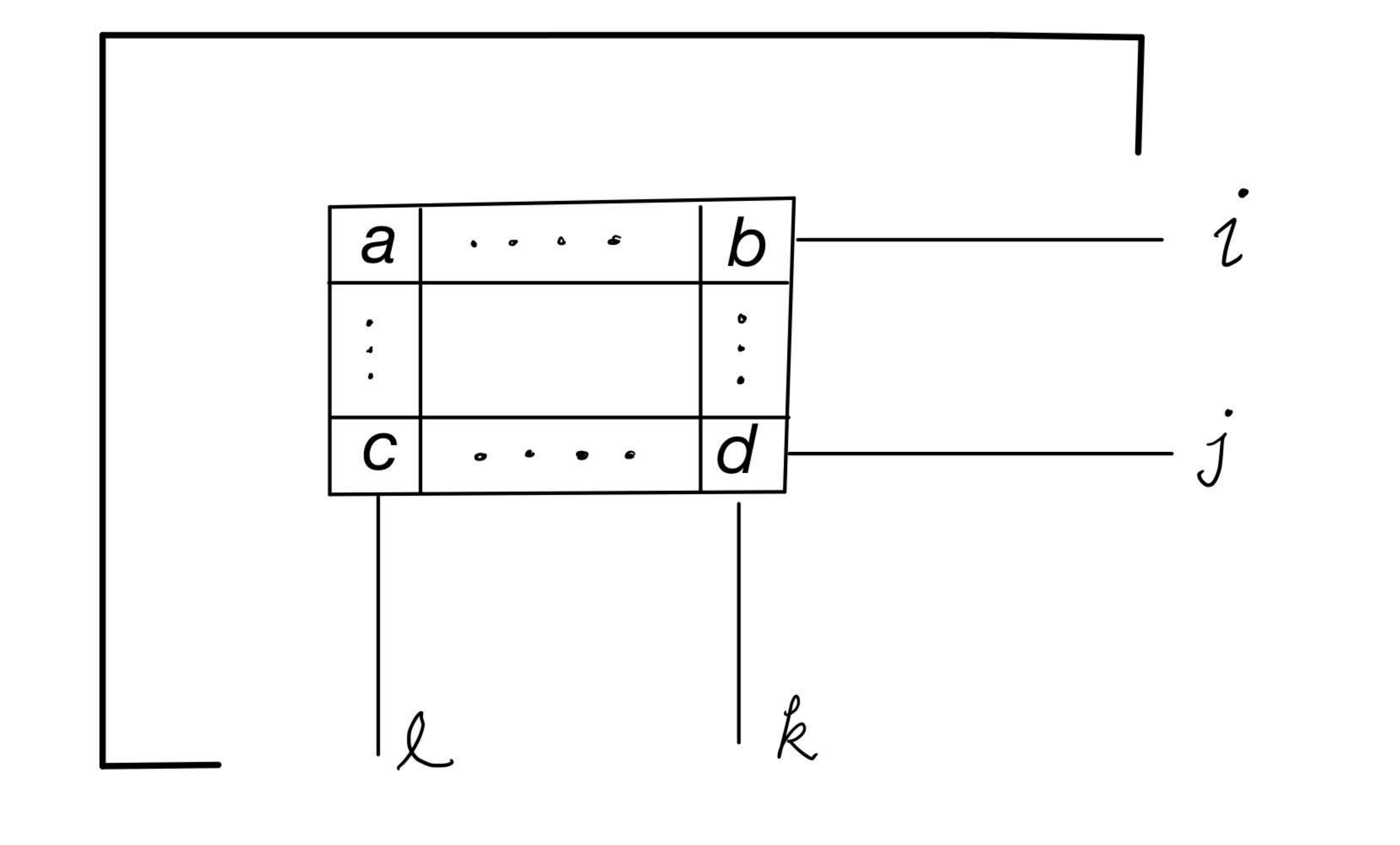} \hskip1cm \raisebox{2cm}{$\begin{array}{ll}a=\ell([i,l])\\
b=\ell([i,k])\\
c=\ell([j,l])\\
d=\ell([j,k])\end{array}$}
\end{figure}

\begin{itemize}
\item[\rm{(i)}] $C_{a,b}=C_{a,c}=C_{cd}=C_{b,d}=0$, and
$C_{a,d}>0, C_{b,c}<0$,
\item[\rm{(ii)}] if one of the corner boxes is empty (no numbered) or the box \young[1][7][$d$] is outside of the $O\Le$-diagram, then either $C_{a,d}>0$ or
$C_{b,c}>0$.
\end{itemize}
\end{proposition}
\proof
Note that in the $\Le$-diagram, the indices $\{i,j\}$ are pivots, and $\{k,l\}$ are non-pivots. Also we have 
$\kappa_i<\kappa_j<\kappa_k<\kappa_l$. Then the proof is just the computation of the coefficients given by the cross ratio \eqref{eq:CR}. For example, the coefficient $C_{a,d}$ is calculated as
\[
C_{a,d}=\frac{(\kappa_i-\kappa_j)(\kappa_l-\kappa_k)}{(\kappa_i-\kappa_k)(\kappa_j-\kappa_l)}>0
\]
It is also easy to show that for the case where \young[1][8][$d$]is outside the diagram (i.e. there is no $d$), we have $C_{b,c}>0$ (in this case note that $\kappa_i<\kappa_k<\kappa_j<\kappa_l$).
\endproof

\begin{example}\label{ex:C59}
Consider Example \ref{ex:P1}. Only the following six coefficients are negative
\[
C_{2,3}, ~C_{2,7}, ~C_{2,9}, ~C_{4,7}, ~C_{4,9}, ~C_{8,9}~<~0.
\]
All other coefficients for $1\le p<q\le 10$ are $C_{p,q}\ge0$ .
\end{example}


\section{The $\tau$-function as the $M$-theta function}\label{sec:Mtheta}
The $\tau$-function \eqref{eq:tauPl} can be expressed as
\begin{align}\label{eq:tauNormal}
\tau(x,y,t)&=\sum_{n=0}^N\sum_{J\in\mathcal{M}_n(A)}\Delta_J(A)E_J(x,y,t)\\
&=\Delta_{I_0}(A)E_{I_0}(x,y,t)\left(1+\sum_{n=1}^N\sum_{J\in\mathcal{M}_n(A)}\frac{\Delta_J(A)E_J}{\Delta_{I_0}(A)E_{I_0}}\right)\nonumber
\end{align}
Since the solution is given by the second derivative of $\ln\tau$, one can take the $\tau$-function in the following form,
\begin{equation}\label{eq:grammian}
\tau(x,y,t)=1+\sum_{n=1}^N\sum_{J\in\mathcal{M}_n(A)}\Delta_J(A)\frac{E_J(x,y,t)}{E_{I_0}(x,y,t)}.
\end{equation}
where we have used $\Delta_{I_0}(A)=1$ for the pivot set $I_0$.

Then the following theorem is proven in \cite{K:24}.
\begin{theorem}\label{thm:theta-tau}
Given irreducible $A\in\text{Gr}(N,M)_{\ge0}$, the $\tau$-function \eqref{eq:grammian}
is the $M$-theta function \eqref{eq:Theta}, i.e.
\begin{align}\label{eq:Mtheta}
\tau(x,y,t)&=\tilde{\vartheta}_g(\z;\widetilde\Omega)=\sum_{m\in\{0,1\}^{g}}\exp 2\pi i\left(\sum_{p<q}m_pm_q\tilde\Omega_{p,q}+\sum_{p=1}^{g}m_pz_p\right)\\
&=1+\sum_{p=1}^g e^{\widetilde\phi_p}+\sum_{p<q}C_{p,q}e^{\tilde\phi_p+\tilde\phi_q}+\cdots +\left(\prod_{p<q}C_{p,q}\right)e^{\sum_{p=1}^g \tilde\phi_p},
\end{align}
where $g=|P_1(A)|$ and  $2\pi i z_p=\tilde\phi_p(x,y,t)=\phi_p(x,y,t)+\phi^0_p$ for
$p=\ell([i_k,j^{(k)}_m])$ (i.e. $\alpha_p=\kappa_{i_k}, \beta_p=\kappa_{j_m^{(k)}}$) with the ordering $\ell$ in $P_1(A)$ and arbitrary constants $\phi_p^0\in\R$,
\begin{align*}
&\phi_{p}=\xi_{j^{(k)}_m}-\xi_{i_k}=(\beta_p-\alpha_p)\,x+(\beta_p^2-\alpha_p^2)\,y+(\beta_p^3-\alpha_p^3)\,t,\\
&e^{\phi^0_{p}}= A_{k,j_m^{(k)}}\prod_{q\ne p}\frac{\alpha_q-\beta_p}
{\alpha_q-\alpha_p},\\
&C_{p,q}=\exp\left(2\pi i\,\tilde\Omega_{p,q}\right)=\frac{(\alpha_p-\alpha_q)(\beta_p-\beta_q)}{(\alpha_p-\beta_q)(\beta_p-\alpha_q)}.
\end{align*}
Here $q=\ell([i_l,j_n^{(l)}])$, and $A_{k,j^{(k)}_m}$ is the entry in $A$ corresponding to the element
$[i_k,j_m^{(k)}]\in P_1(A)$.
\end{theorem}

As shown in \cite{HKL:24}, the sign of $A_{k,j_m^{(k)}}$ is determined by the positivity of $e^{\phi^0_p}$, that is, it is the sign of the product, $\prod_{q\ne p}\frac{\alpha_q-\beta_p}{\alpha_q-\alpha_p}$, in the equation. 

\begin{remark}\label{rem:Hirota}
One should note that the $\tau$-function \eqref{eq:Mtheta} in Theorem \ref{thm:theta-tau}
can be used for the general $N\times M$ matrix $A$ in RREF. In particular, if there is only one nonzero
entry in each row besides the pivot, we have $g=N$ and $A\in\Gr(N,2N)$. Then the solution $u(x,y,t)$
generated by the $\tau$-function gives $N$-soliton solution of Hirota-type (just $N$ line solitons without resonances).
In general, this soliton solution in \emph{singular}, except if $A\in\Gr(N,M)_{\ge 0}$.
\end{remark}

\subsection{Example}\label{sec:Ex24}
Consider the $O\Le$-diagram \young[2,2][6][1,2,3,4]. This implies $g=4$, and the number in each box of the diagram is assigned by $p=\ell([i,j])$ for $[i,j]\in P_1(A)$ with $A\in\text{Gr}(2,4)_{\ge 0}$, i.e.
\begin{equation}\label{eq:lij}
1=\ell([1,4]),\quad 2=\ell([1,3]),\quad 3=\ell([2,4]), \quad 4=\ell([2,3]).
\end{equation}
That is, we have the following table,
\begin{table}[h]
			\centering
			\begin{tabular}{|c||c|c|c|c|}
				\hline
				$p$ & $1$ & $2$ &$3$ &$4$  \\  
				\hline 
				$\alpha_p$   & $\kappa_{1}$ & $\kappa_{1}$ & $\kappa_{2}$ & $\kappa_{2}$  \\ 
				\hline  
				$\beta_p$ & $\kappa_{4}$ & $\kappa_{3}$  & $\kappa_{4}$ & $\kappa_{3}$  \\   
				\hline
			\end{tabular}
		\end{table}
		
In terms of the normalization \eqref{eq:normal}, 
this ordering means $\pi^{-1}(s_p)=\{\alpha_p,\beta_p\}$ for $p=1,\ldots,4$, e.g., $\pi^{-1}(s_2)=\{\kappa_1,\kappa_3\}$
(see \eqref{eq:NOP1}).
Then the coefficients $C_{p,q}$ in \eqref{eq:CR} are calculated as $C_{1,2}=C_{1,3}=C_{2,4}=C_{3,4}=0$, and
\[
C_{1,4}=\frac{(\kappa_1-\kappa_2)(\kappa_4-\kappa_3)}{(\kappa_1-\kappa_3)(\kappa_4-\kappa_2)}>0,\qquad
C_{2,3}=\frac{(\kappa_1-\kappa_2)(\kappa_3-\kappa_4)}{(\kappa_1-\kappa_4)(\kappa_3-\kappa_2)}<0.
\]
The matrix $A\in\text{Gr}(2,4)_{\ge0}$ corresponding to the diagram is given by
\[
A=\begin{pmatrix}
1 & 0 & A_{1,3} & A_{1,4}\\
0&1& A_{2,3}& A_{2,4}
\end{pmatrix}.
\]
The signs of the entries $A_{i,j}$ are determined by the positivity of $\exp\phi_p^0$ (since $\phi_p^0\in\R$),
\begin{align*}
e^{\phi_1^0}=A_{1,4}\frac{\kappa_2-\kappa_4}{\kappa_2-\kappa_1}>0,\qquad &e^{\phi_2^0}=A_{1,3}\frac{\kappa_2-\kappa_3}{\kappa_2-\kappa_1}>0,\\
e^{\phi_3^0}=A_{2,4}\frac{\kappa_1-\kappa_4}{\kappa_1-\kappa_2}>0,\qquad &e^{\phi_4^0}=A_{2,3}\frac{\kappa_1-\kappa_3}{\kappa_1-\kappa_2}>0,
\end{align*}
that is, using $\kappa_1<\kappa_2<\kappa_3<\kappa_4$, we have $A_{1,4}, A_{1,3}<0$ and $ A_{2,4},A_{2,3}>0$. Note here that these signs are \emph{not} enough for the total nonnegativity of $A$ (the additional condition is determined by the regularity of the solution \cite{KW:13}, see below).

Then the $M$-theta function (i.e. the $\tau$-function) in Theorem \ref{thm:theta-tau} is given by
\begin{equation}\label{eq:tau24}
\tau=1+e^{\widetilde\phi_1}+e^{\widetilde\phi_2}+e^{\widetilde\phi_3}+e^{\widetilde\phi_4}+C_{1,4}e^{\widetilde\phi_1+\widetilde\phi_4}+C_{2,3}e^{\widetilde\phi_2+\widetilde\phi_3},
\end{equation}
where the exponents are given by $\widetilde\phi_p=\phi_p+\phi_p^0$ with $\phi_p=\xi_j(x,y,t)-\xi_i(x,y,t)$  in 
\eqref{eq:E} for $p=\ell([i,j])=1,\ldots,4$,
\[
\phi_1=\xi_4-\xi_1,\quad \phi_2=\xi_3-\xi_1,\quad \phi_3=\xi_4-\xi_2,\quad\phi_4=\xi_3-\xi_2,
\]
One should note here that we have a linear relation among the phase functions $\phi_i$'s, i.e.
\[
\phi_1+\phi_4=\phi_2+\phi_3=(\xi_3+\xi_4)-(\xi_1+\xi_2).
\]
Then the last two terms in the $\tau$-function \eqref{eq:tau24} becomes
\begin{align*}
\left(C_{1,4}e^{\phi_1^0+\phi_4^0}+C_{2,3}e^{\phi_2^0+\phi_3^0}\right)e^{\phi_1+\phi_4}=
(A_{1,3}A_{2,4}-A_{1,4}A_{2,3})\frac{\kappa_3-\kappa_4}{\kappa_1-\kappa_2}e^{\phi_1+\phi_4}.
\end{align*}
This implies that for the regular soliton solution, we need to choose appropriate constants
$\phi_1^0,\ldots,\phi_4^0$ so that  $A_{1,3}A_{2,4}-A_{14}A_{2,3}\ge 0$, i.e. $A\in\text{Gr}(2,4)_{\ge0}$.


\section{The Schottky uniformization}\label{sec:Schottky}
A goal of the present paper is to construct a smooth compact Riemann surface $\mathcal{R}_g$
associated with the KP soliton whose $M$-theta function $\widetilde\vartheta_g$ is obtained by taking a tropical (singular) limit of $\mathcal{R}_g$. We achieve the goal using the Schottky uniformization theorem \cite{Ford, Ba:97, Bur1:91, Bur2:91}.
A Schottky group is defined as a finitely generated, discontinuous subgroup of $PSL_2(\C)$ which are free and purely loxodromic \cite{Ford, Ba:97, BBEIM:94}. In this paper, we consider
a special case of the Schottky group, which is generated by purely hyperbolic M\"obius transformations in $PSL_2(\R)$.
It is well-known after Poincar\'e and Koebe \cite{Ford} that any compact Riemann surface $\RR$ can be uniformized by the Schottky group $\Gamma$, which can be represented as
\[
\RR~\cong~\Omega(\Gamma)/\Gamma,
\]
where $\Omega(\Gamma)$ is the set of discontinuity of $\Gamma$ (see also \cite{BBEIM:94}).

In order to define our Schottky group $\Gamma_A$ for $A\in\text{Gr}(N,M)_{\ge0}$, we start with the following definition.
\begin{definition}\label{def:orderij}
For each element $[i,j]\in P_1(A)$, we define a pair of real numbers $\{\kappa_{i,j},\kappa_{j,i}\}$
with the order,
\begin{itemize}
\item[\rm{(a)}] $\kappa_k<\kappa_{k,\bullet}<\kappa_l<\kappa_{l,\bullet}$ for all $k<l\in[M]$, and
\item[\rm{(b)}] $\kappa_{k,p}<\kappa_{k,q}$, when $p>q$ and for $k\in[M]$.
\end{itemize}
We write $\alpha_p=\kappa_{i,j}, \beta_p=\kappa_{j,i}$ with $p=\ell([i,j])$, the ordering $\ell$ in $P_1(A)$.
\end{definition}
Let $\gamma_p$ be a hyperbolic M\"obius transform on $\C\mathbb{P}^1$ having two fixed points $\{\alpha_p,\beta_p\}$, which is given by
\begin{equation}\label{eq:FLT}
\frac{\gamma_p(z)-\alpha_p}{\gamma_p(z)-\beta_p}
=\mu_p\frac{z-\alpha_p}{z-\beta_p}, 
\end{equation}
where $\mu_p$ is the multiplier which is a real constant with $0<\mu_p<1$. 
Then the fixed points $\alpha_p$ and $\beta_p$ are attractive and repulsive, respectively.
Then we define the Schottky group $\Gamma_A$ associated with $A\in\text{Gr}(N,M)_{\ge0}$ as
a Fuchsian group given by
\begin{equation}\label{eq:SGroup}
\Gamma_A:=\langle~ \gamma_p\in {PSL}_2(\R): p=\ell([i,j]) ~\text{for}~[i,j]\in P_1(A)~\rangle.
\end{equation}
where $\gamma_p$ in \eqref{eq:FLT} is expressed as
\begin{equation}\label{eq:gamma}
\gamma_p=\frac{1}{(\beta_p-\alpha_p)\sqrt{\mu_p}}
\begin{pmatrix}
\alpha_p-\mu_p\beta_p & -\alpha_p\beta_p(1-\mu_p)\\
1-\mu_p & -(\beta_p-\mu_p\alpha_p)
\end{pmatrix}.
\end{equation}
In Section \ref{sec:deformation} below, 
we directly construct $\gamma_p$ as a deformation of the singular curve (Riemann surface) 
associated with each element $A\in\text{Gr}(N,M)_{\ge0}$.

The isometric circle $I(\gamma_p)$ of $\gamma_p$ in \eqref{eq:gamma} is defined by $|d\gamma_p(z)/dz|=1$ \cite{Ford}, which gives
\[
|(1-\mu_p)z-(\beta_p-\mu_p\alpha_p)|=(\beta_p-\alpha_p)\sqrt{\mu_p},
\]
whose center and radius are 
\begin{equation}\label{eq:C-R}
\text{Center}=\frac{\beta_p-\mu_p\alpha_p}{1-\mu_p},\qquad
\text{Radius}=\frac{\beta_p-\alpha_p}{1-\mu_p}\sqrt{\mu_p}.
\end{equation}
Taking $\mu_p$ small enough, one can assume that all the isometric circles are disjoint.
Note that $\gamma_p$ maps outside of the isometric circle $I(\gamma_p)$ into
the interior of $I(\gamma_p^{-1})$, see the figure below.

\begin{figure}[h]
\centering
\includegraphics[width=9cm]{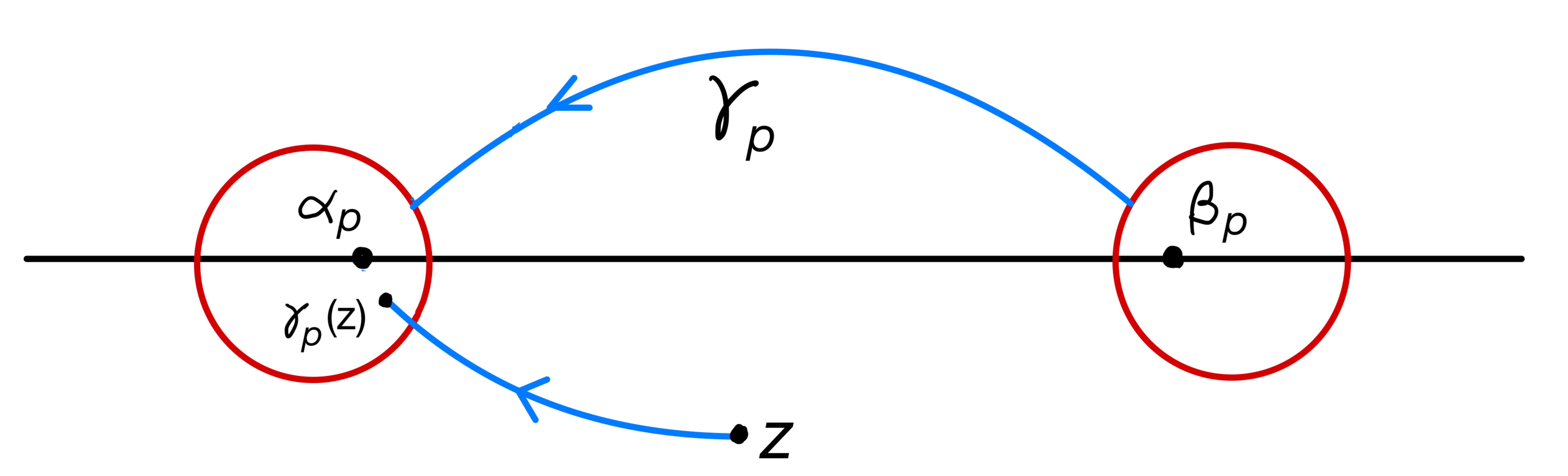}
\end{figure}

\noindent
The (isometric) fundamental region, denoted by $\mathcal{F}(\Gamma_A)$, of $\Gamma_A$ is given by $\C\mathbb{P}^1$ with $2g$ holes of isometric circles, i.e.
\begin{equation}\label{eq:F}
\mathcal{F}(\Gamma_A):=\text{Ext}\,\left({\bigcup_{p=1}^g\,\overline{\text{Int}\,\left(I(\gamma_p)\right)}\cup \text{Int}\,\left(I(\gamma_p^{-1})\right)}\right),
\end{equation}
where $\text{Ext}({D})$ means the set of exterior points of the set ${D}$, and
$\text{Int}(I(\gamma))$ represents the interior points of the isometric circle $I(\gamma)$.

For each $[i,j]\in P_1(A)$, let $\omega_p$  with $p=\ell([i,j])$ be the differentials on $\Omega(\Gamma_A)$, the set of discontinuity of $\Gamma_A$, defined by
\begin{equation}\label{eq:Uomega}
\omega_p(z)=\frac{dz}{2\pi i}\sum_{\gamma\in\Gamma_A/\langle\gamma_p\rangle}
\left(\frac{1}{z-\gamma(\alpha_p)}-\frac{1}{z-\gamma(\beta_p)}\right)\qquad \text{on}\quad \Omega(\Gamma_A),
\end{equation}
where $\gamma$ runs through all representatives of the right coset classes of $\Gamma_A$ by its cyclic subgroup $\langle\gamma_p\rangle$ generated by $\gamma_p$. Here $\Omega(\Gamma_A)$ can be expressed as $\Omega(\Gamma_A)=\cup_{\gamma\in\Gamma_A}\gamma(\mathcal{F}(\Gamma_A))$. 
It is also known \cite{Ford, Ba:97, Sch:13} that the infinite sum in \eqref{eq:Uomega} converges absolutely for sufficiently small $\mu_p$. Then we have the lemma.
\begin{lemma}\label{lem:holomorphic}
The differentials $\omega_p$ are holomorphic on $\Omega(\Gamma_A)$, 
\[
\omega_p(z)=\omega_p(\gamma(z))\qquad\text{for any}\quad \gamma\in\Gamma_A.
\]
\end{lemma}
\proof
Let $\omega$ be a differential given by
\[
\omega(z)=\left(\frac{1}{z-\alpha}-\frac{1}{z-\beta}\right)\,dz=\frac{\alpha-\beta}{(z-\alpha)(z-\beta)} \,dz.
\]
Then for $\sigma\in\Gamma_A$, we have
\[
\omega(\sigma(z))=\frac{\sigma^{-1}(\alpha)-\sigma^{-1}(\beta)}{(z-\sigma^{-1}(\alpha))(z-\sigma^{-1}(\beta))}\,dz.
\]
Then taking $\alpha=\gamma(\alpha_p)$ and $\beta=\gamma(\beta_p)$, and then
$\sigma^{-1}\gamma\in\Gamma_A/\langle\gamma_p\rangle$. Summing over
all the element in $\Gamma_A/\langle\gamma_p\rangle$ gives a proof.
\endproof

For each $[i,j] \in P_{1}(A)$ with $p=\ell([i,j])$, 
we define the the associated $a$-cycle $a_p$ 
as the isometric circle $I(\gamma_p)$ with counter-clockwise orientation, 
and the $b$-cycle $b_p$ as an oriented path 
from a point $a$ on $I(\gamma_p) \cap {\mathbb R}$ to 
$\gamma_p(a) \in I(\gamma_p^{-1}) \cap {\mathbb R}$ in $\mathcal{F}(\Gamma_{A})$.
Then we have following proposition \cite{Ba:97, Bur1:91, Bur2:91} (see also \cite[Theorem 3.5]{I:22}). 

\begin{proposition}\label{prop:PeriodI}
For $p=\ell([i,j])$ and $q=\ell([k,l])$, the differential \eqref{eq:Uomega} is normalized, i.e.
\[
\oint_{a_p}\omega_{q}=\delta_{p,q}, 
\]
and the period matrix $\Omega_A$ is calculated as
\begin{equation}
(\Omega_A)_{p,q}=\oint_{b_p}\omega_{q}=\frac{1}{2\pi i}\sum_{\gamma\in\langle\gamma_p\rangle\setminus\Gamma_A/\langle\gamma_{q}\rangle}\ln\,[\alpha_p,\beta_p;\gamma(\alpha_q),\gamma(\beta_q)] \qquad\text{mod}(\Z), \label{eq:Rperiod}
\end{equation}
where $[\alpha_p,\beta_p;\gamma(\alpha_q),\gamma(\beta_q)]$ is the cross ratio given by
\[
[\alpha_p,\beta_p;\gamma(\alpha_q),\gamma(\beta_q)]:=\frac{(\alpha_p-\gamma(\alpha_q))(\beta_p-\gamma(\beta_q))}{(\alpha_p-\gamma(\beta_q))(\beta_p-\gamma(\alpha_q))},
\]
which takes $\mu_p$ when $p=q$ and $\gamma\in\langle\gamma_p\rangle$. 
Note here that the sign of the nonzero real number $\exp \left( 2 \pi i \oint_{b_p} \omega_q \right)$ 
is equal to that of $C_{p,q} = [\alpha_p, \beta_p; \alpha_q, \beta_q]$. 
\end{proposition}
\proof
The period integral over $a_p$ are obvious, and this implies that $\omega_p$ is normalized. 
By \eqref{eq:Uomega}, 
\begin{align*}
\exp \left( 2 \pi i \oint_{b_p}\omega_{q} \right)&=
\prod_{\gamma\in\Gamma_A/\langle\gamma_{q}\rangle} 
\left.\frac{z-\gamma(\alpha_q)}{z-\gamma(\beta_q)}\right|_a^{\gamma_p(a)}\\
&= \prod_{\gamma\in\Gamma_A/\langle\gamma_{q}\rangle} 
\frac{(\gamma_p(a)-\gamma(\alpha_q))(a-\gamma(\beta_q))}
{(\gamma_p(a)-\gamma(\beta_q))(a-\gamma(\alpha_q))}.
\end{align*}
Here, if $p=q$ and $\gamma \in \langle \gamma_p \rangle$, 
then by (6.1), 
\begin{align*}
\frac{(\gamma_p(a)-\gamma(\alpha_q))(a-\gamma(\beta_q))}{(\gamma_p(a)-\gamma(\beta_q))(a-\gamma(\alpha_q))} 
= 
\frac{(\gamma_p(a) - \alpha_p)(a - \beta_p)}
{(\gamma_p(a) - \beta_p)(a - \alpha_p)} 
= \mu_{p}. 
\end{align*}
Since $\lim_{n \rightarrow \infty} \gamma_{p}^{n}(a) = \alpha_p$ and 
$\lim_{n \rightarrow \infty} \gamma_{p}^{-n}(a) = \beta_p$, 
if $p \neq q$ or $\gamma \not\in \langle \gamma_p \rangle$, 
then 
\begin{eqnarray*}
\lefteqn{
\prod_{\gamma \in \Gamma_A/\langle \gamma_{[k,l]} \rangle} 
\frac{(\gamma_p(a) - \gamma(\alpha_q))(a - \gamma(\beta_q))}
{(\gamma_p(a) - \gamma(\beta_q))(a - \gamma(\alpha_q))} 
} 
\\ 
& = & 
\prod_{\gamma \in \langle\gamma_p\rangle\setminus \Gamma_A/\langle \gamma_{q} \rangle} 
\prod_{n \in {\mathbb Z}} 
\frac{(\gamma_p(a) - \gamma_p^{-n} \gamma(\alpha_q))
(a - \gamma_p^{-n} \gamma(\beta_q))}
{(\gamma_p(a) - \gamma_p^{-n} \gamma(\beta_q))
(a - \gamma_p^{-n}\gamma(\alpha_q))} 
\\ 
& = & 
\prod_{\gamma \in \langle\gamma_p\rangle\setminus \Gamma_A/\langle \gamma_{q} \rangle} 
\prod_{n \in {\mathbb Z}} 
\frac{(\gamma_p^{n+1}(a) - \gamma(\alpha_q))
(\gamma_p^{n}(a) - \gamma(\beta_q))}
{(\gamma_p^{n+1}(a) - \gamma(\beta_q))
(\gamma_p^{n}(a) - \gamma(\alpha_q))} 
\\ 
& = & 
\prod_{\gamma \in \langle\gamma_p\rangle\setminus \Gamma_A/\langle \gamma_{q} \rangle} 
\frac{(\alpha_p - \gamma(\alpha_q))(\beta_p - \gamma(\beta_q))}
{(\alpha_p - \gamma(\beta_q))(\beta_{p} - \gamma(\alpha_q))}. 
\end{eqnarray*}
In the above infinite product, 
if 
$\gamma \not\in \langle \gamma_{p} \rangle \cdot \langle \gamma_{p} \rangle$, 
then $\gamma(\alpha_q)$ and $\gamma(\beta_q)$ belong to the interior of 
$I(\gamma_r)$ or $I(\gamma_r^{-1})$ for $r \neq p$ which implies that 
$[\alpha_p,\beta_p;\gamma(\alpha_q),\gamma(\beta_q)] > 0$. 
This completes the proof. 
\endproof
We remark that in the limits $\mu_p\to 0$ for all $p=1,\ldots,g$, we obtain \eqref{eq:CR}, i.e.
\[
\exp\left(2\pi i\,\Omega_{p,q}\right)\quad \longrightarrow\quad C_{p,q}=[\alpha_p,\beta_p;\alpha_q,\beta_q].
\]

As a summary of these results, we now give the main theorem of the paper.
\begin{theorem} \label{thm:main}
Given irreducible $A\in\text{Gr}(N,M)_{\ge 0}$, a real compact Riemann surface $\mathcal{R}_g$
can be constructed by the Schottky group $\Gamma_A$ defined in \eqref{eq:SGroup} with \eqref{eq:gamma},
i.e.
\[
\mathcal{R}_g\cong \Omega(\Gamma_A)/\Gamma_A,
\]
where $g=|P_1(A)|$ in \eqref{eq:P1} and $\Omega(\Gamma_A)$ is the set of discontinuity of $\Gamma_A$. The $\vartheta$-function  defined on $\mathcal{R}_g$  is 
given by \eqref{eq:Riemann} with the period matrix in \eqref{eq:Rperiod}.
\end{theorem}


\subsection{Deformation of nodal singular surfaces}\label{sec:deformation} 
By the limiting process 
$\kappa_{k, \bullet} \rightarrow \kappa_{k}$ and $\mu_{p} \rightarrow 0$, 
the Riemann surface $\Omega(\Gamma_A)/\Gamma_A$ tends to 
the singular Riemann surface whose singular points are not nodal in general. 
In this section, we explain how we construct the Schottky uniformized deformation of 
nodal singular curves associated with TNN Grassmannians.

Let us first define an oriented graph $\Delta_A(V,E)$ associated with the element $A\in\text{Gr}(N,M)_{\ge 0}$, whose sets of vertices $V$ and oriented edges $E$ are given as follows:
\begin{itemize}
\item[(a)] $V:=\{v_0, v_k\,(k\in [M])\}$,
\item[(b)] $E:=\{e_k\,(k\in [M]),\,\, e_{[i,j]}\,([i,j]\in P_1(A))\}$,
\end{itemize}
where each edge $e_k$ is from $v_0$ to $v_k$, and $e_{[i,j]}$ from $v_i$ to $v_j$. Then the set of closed paths $e_i\cdot e_{[i,j]}\cdot e_j^{-1}$ forms the fundamental group $\pi_1(\Delta_A,v_0)$ with the base point
$v_0$. The homological group $H_1(\Delta_A;\Z)$ is then given by Abelianization of the group $\pi_1$ and the dimension is
$\text{dim}H_1(\Delta,\Z)=|P_1(A)|$. 

We call algebraic curves defined over ${\mathbb R}$ {\it real curves}, 
and construct a singular real curve $\mathcal{C}_{A}$ with dual graph $\Delta_{A}$ 
and a family of real curves ${\mathcal R}_{A}$ as the deformation of $\mathcal{C}_{A}$. 
Denote by $\R\mathbb {P}^{1}$ the real projective line 
${\mathbb R} \cup \{ \infty \}$ which is identified with an oriented circle 
according to the increase of real numbers. 
Put $\mathcal{P}_{v_{0}} = \R\mathbb{P}^{1}$ with counter-clockwise orientation, 
and take points $\kappa_{k}$ $(k \in [M])$ on $\mathcal{P}_{v_{0}} \setminus \{ \infty \}$ 
with the ordering \eqref{eq:orderK}.

For each vertex $v_{k}$ $(k \in [M])$, 
put $\mathcal{P}_{v_{k}} = \R\mathbb{P}^{1}$ with counter-clockwise orientation, 
and take points $\lambda_{k} \in \mathcal{P}_{v_{k}} \setminus\{ \infty \}$ and 
$\lambda_{k, l} \in \mathcal{P}_{v_{k}} \setminus\{ \infty \}$ if $[k, l] \in P_{1}(A)$ or $[l, k] \in P_{1}(A)$
such that $\lambda_{k,l} < \lambda_{k}$ and $\lambda_{k, l} < \lambda_{k, m}$ for $l > m$.
Then the singular real curve $\mathcal{C}_{A}$ with dual graph $\Delta_{A}$ is obtained 
as a union of $\mathcal{P}_{v_{0}}$ and $\mathcal{P}_{v_{k}}$ $(k \in [M])$ by identifying 
$$
\kappa_{k} = \lambda_{k} \ (k \in [M]), \qquad
\lambda_{i, j} = \lambda_{j, i} \ ([i, j] \in P_{1}(A)). 
$$
Note that the singular points on $\mathcal{C}_{A}$ are nodal, 
and that the (arithmetic) genus of $\mathcal{C}_{A}$ is $g=|P_{1}(A)|$. 
For small positive parameters $\nu_{k}$ $(k \in [M])$ and $\nu_{i, j} = \nu_{j, i}$ $([i, j] \in P_{1}(A))$, 
let ${\mathcal R}_{A}$ be a family of real curves as the deformation of $\mathcal{C}_{A}$ 
obtained by gluing 
$$
\mathcal{C}_{A} \setminus\{ \mbox{neighborhoods of singular points} \}
$$ 
under the relations 
\begin{equation}\label{eq:Re1}
(z_{0} - \kappa_{k})(z_{k} - \lambda_{k}) = -\nu_{k}, 
\end{equation}
and 
\begin{equation}\label{eq:Re2}
(z_{i} - \lambda_{i, j})(z_{j} - \lambda_{j, i}) = -\nu_{i, j}, 
\end{equation} 
where $z_{i}$ are the coordinates of $\mathcal{P}_{v_{i}}$. 
By these relations, for $[i, j] \in P_{1}(A)$, 
if $z, w \in \mathcal{P}_{v_{0}} = \R\mathbb{P}^{1}$ are related as
$$ 
z \in \mathcal{P}_{v_0} \quad \stackrel{\eqref{eq:Re1}}{\longmapsto} \quad 
z_{i} \in\mathcal{P}_{v_i} \quad \stackrel{\eqref{eq:Re2}}{\longmapsto} \quad  
z_{j} \in\mathcal{P}_{v_p} \quad \stackrel{\eqref{eq:Re1}}{\longmapsto} \quad w\in\mathcal{P}_{v_0}, 
$$
then we have
\begin{align*}
w - \kappa_{j} &= -\frac{\nu_j}{z_j-\lambda_j}=\frac{a\nu_j(z-\kappa_i)-\nu_i\nu_j}{(ab+\nu_{i,j})(z-\kappa_i)-b\nu_i}
\end{align*}
where $a = \lambda_{i} - \lambda_{i,j}$ and $b = \lambda_{j} - \lambda_{j,i}$. This gives the M\"obius transform $\gamma:z\mapsto w={\gamma}(z)$ on $\mathcal{P}_{v_0}$ with $\gamma\in {PSL}_2(\R)$,
\[
\gamma=\frac{1}{\sqrt{\nu_i\nu_j\nu_{i,j}}}
\begin{pmatrix}
c\kappa_j +a\nu_j & -c\kappa_i\kappa_j-\nu_i\nu_j-a\kappa_i\nu_j-b\kappa_j\nu_i\\
c & -c\kappa_i-b\nu_i
\end{pmatrix},
\]
where $c=ab+\nu_{i,j}$. 
Then introducing the Schottky parameters 
$\{\alpha_p=\kappa_{i,j},\beta_p=\kappa_{j,i},\mu_p\}$ in terms of $\{a\nu_j,b\nu_i,c\}$, 
we have $\gamma_p$ defined in \eqref{eq:gamma}.
We can also see 
$$
\kappa_{k,l} - \kappa_{k} = \Theta(\nu_{k}), 
\qquad \mu_{i, j} = \Theta(\nu_{i} \nu_{i,j} \nu_{j}),
$$
where $f = \Theta(g)$ means that there exists positive constants $c_{1}, c_{2}$ 
satisfying $c_{1} |g| \leq |f| \leq c_{2} |g|$ asymptotically. 
Therefore, 
as is shown in \cite[Section 3]{I:00}, 
${\mathcal R}_{A}$ with sufficiently small $\nu_{k}, \nu_{i, j} > 0$ 
gives a family of real curves which are Schottky uniformized by real Schottky groups 
$\Gamma_{A}$ with free generators $\gamma_{[i, j]}$ $([i, j] \in P_{1}(A))$. 
Furthermore, 
under $\nu_{k}, \nu_{i, j} \rightarrow 0$, 
we have $\alpha_p=\kappa_{i,j} \rightarrow \kappa_{i}$, $\beta_p=\kappa_{j,i} \rightarrow \kappa_{j}$ and 
$\gamma(\alpha_p) - \gamma(\beta_p) \rightarrow 0$ 
for any 
$\gamma \in (\Gamma_{A} \setminus \langle \gamma_{p} \rangle)/\langle \gamma_{p} \rangle$. 
Therefore, 
the differentials $\omega_{p}$ with $p=\ell([i,j])$ given in \eqref{eq:Uomega} has the limit
$$
\omega_{p}\quad\longrightarrow\quad \frac{dz}{2 \pi i} \left( \frac{1}{z - \kappa_{i}} - \frac{1}{z - \kappa_{j}} \right), 
$$
and by Proposition \ref{prop:PeriodI}, the period matrix $\Omega_{p,q}$ with $p=\ell([i,j])$ and $q=\ell([k,l])$ has the limit
$$
\exp \left( 2 \pi i \oint_{b_{p}} \omega_{q} \right)~ \longrightarrow ~
\left\{ \begin{array}{ll} 
0 & (i = k \ \mbox{or} \ j = l), 
\\
\left[ \kappa_{i}, \kappa_{j}; \kappa_{k}, \kappa_{l} \right] & (i \neq k \ \mbox{and} \ j \neq l). 
\end{array} \right.  
$$   
Taking appropriate pairs $\{\alpha_p,\beta_p\}$ in the normalization in Section \ref{sec:SRiemann}, we recover the limits in  \eqref{eq:Lomega} and \eqref{eq:LPmatrix}.

\begin{remark}
We can give a sufficient condition that the right hand side of \eqref{eq:Rperiod} 
is convergent. 
Under fixing parameters $\{\lambda_{k}, \nu_{k}, \lambda_{i,j}=\lambda_{j,i}\}$, 
the above calculation and results of \cite[Section 3]{I:00}, \cite[Section 3]{I:22} leads to the formula \eqref{eq:Rperiod} in Proposition \ref{prop:PeriodI},
\begin{equation}\label{eq:Mperiod}
\exp \left( 2 \pi i \oint_{b_{p}} \omega_{q} \right) = 
\prod_{\gamma\in\langle\gamma_{p}\rangle\setminus\Gamma_A/\langle\gamma_{q}\rangle}[\alpha_p,\beta_p;\gamma(\alpha_q),\gamma(\beta_q)],
\end{equation}
which can be expressed as formal power series in $\nu_{i,j}$ $([i,j] \in P_{1}(A))$.
The series is convergent, if $\nu_{i,j}$ are sufficiently small. 
Let $r_{i,j}$ $([i,j] \in P_{1}(A))$ be positive numbers such that 
for any complex numbers $\nu_{i,j}$ with $|\nu_{i,j}| < r_{i,j}$, 
by the relations \eqref{eq:Re1}, \eqref{eq:Re2}, 
one obtain smooth Riemann surfaces. 
Then the left hand side of \eqref{eq:Mperiod} is a holomorphic function of 
$\nu_{i,j}$ $([i,j] \in P_{1}(A))$ with $|\nu_{i,j}| < r_{i,j}$, 
and hence the right hand side is convergent for these $\nu_{i,j}$'s (see also \cite{Sch:13}).
\end{remark}

\subsection{Quasi-periodic solutions}
In this section, we just recall the formula in \cite{BB:89} which gives a quasi-periodic solution in terms of the theta function \eqref{eq:Riemann} using the Schottky group. The explicit formula of the solution $u(x,y,t)$ is given by
\begin{equation}\label{eq:QPS}
u(x,y,t)=2\,\partial_x^2 \,\ln\,\vartheta_g({\bf U}^1x+{\bf U}^2y+{\bf{U}}^3t+{\bf{D}};\, \Omega_A)+2C
\end{equation}
where ${\bf U}^k=(U^k_{p}: 1\le p\le g)$ for $k=1,2,3$ are $g$-dimensional vectors with $g=|P_1(A)|$ given by
\begin{align*}
U_{p}^k&:=\frac{1}{2\pi i}\,\sum_{\gamma\in\Gamma_A/\langle\gamma_{p}\rangle}\left(\gamma(\beta_p)^k-\gamma(\alpha_p)^k\right).
\end{align*}
The period matrix $\Omega_A$ is given by \eqref{eq:Rperiod}, and  $2\pi i{\bf D}\in\R^g$ is an arbitrary constant vector. The constant $C$ is computed as
\[
C=\sum_{p=1}^g\left(\frac{(\beta_p-\alpha_p)\sqrt{\mu_p}}{1-\mu_{p}}\right)^2.
\]
Now it is easy to confirm that the quasi-periodic solution $u(x,y,t)$ given in \eqref{eq:QPS} leads to the KP soliton in the limit with
 $\alpha_p\to\kappa_i$, $\beta_p\to\kappa_j$ and $\mu_p\to0$.

\begin{remark}
In general, our construction of a real compact Riemann surface $\mathcal{R}$ does not give
the so-called $\mathsf{M}$-curve {\rm \cite{DN:89}}, which requires that on $\mathcal{R}$, the involution $\sigma$ must have a maximum number of ovals chosen from the homological basis. Here the involution $\sigma$ acts on $H_1(\mathcal{R};\Z)=\langle a_p, b_p; p=1,\ldots,g\rangle$ by
\[
\sigma(a_p)=a_p,\quad \sigma(b_p)=-b_p,\qquad {\rm for}~\, p=1,\ldots,g.
\] 
In the case that the Riemann surface is not an $\mathsf{M}$-curve, the quasi-periodic solution
of the KP equation is not regular {\rm\cite{DN:89}} (Theorem in p.271). We will discuss in more details in a future communication {\rm\cite{IK:25}}.
\end{remark}

\section{Examples}\label{sec:Example}
Here we give two examples, and illustrate their fundamental domains $\mathcal{F}(\Gamma_A)$ in \eqref{eq:F} for the smooth Riemann surfaces uniformized by the Schottky groups $\Gamma_A$.
We also discuss the regularity of the quasi-periodic solutions based on the regularity of the soliton
solutions in the limits $\mu_p\to0$ in the Schottky group.

\subsection{The cases of $\text{Gr}(2,4)_{\ge0}$}
(a) {\bf The cases with $g=4$:}
Consider the case with the $O\Le$-diagram \young[2,2][6][1,2,3,4]. Then we have
\[
P_1(A)=\{[1,4],\,[1,3],\,[2,4],\,[2,3]\},
\]
and the numbers in the $O\Le$-diagram are $1=\ell([1,4]), 2=\ell([1,3]), 3=\ell([2,4])$ and $4=\ell([2,3])$.
The element $\gamma_p$ with $p=\ell([i,j])$ in the Schottky group $\Gamma_A$ are defined by \eqref{eq:gamma}, where 

\begin{table}[h]
			\centering
			\begin{tabular}{|c||c|c|c|c|}
				\hline
				$p$ & $1$ & $2$ &$3$ &$4$  \\  
				\hline 
				$\alpha_p$   & $\kappa_{1,4}$ & $\kappa_{1,3}$ & $\kappa_{2,4}$ & $\kappa_{2,3}$  \\ 
				\hline  
				$\beta_p$ & $\kappa_{4,1}$ & $\kappa_{3,1}$  & $\kappa_{4,2}$ & $\kappa_{3,2}$  \\   
				\hline
			\end{tabular}
		\end{table}

\noindent
Recall  $\kappa_{k,l}-\kappa_k=\Theta(\nu_k)$ discussed in Section \ref{sec:deformation}, and taking the limit $\nu_k\to0$, we recover the table in Example \ref{sec:Ex24}.

The fundamental domain $\mathcal{F}(\Gamma_A)$ is shown in the figure below, that is, $\mathcal{F}(\Gamma_A)$ is the domain outside the isometric circles given in \eqref{eq:F}.
In the figure, the dots on the real line are $\{\alpha_p,\beta_p:p=1,\ldots,4\}$, and the $b$-cycles are shown as the actions of the group elements $\gamma_p$ for $p=\ell([i,j])$ (i.e. $b_{[i,j]}$ corresponds to $\gamma_p$). Notice that the $b$-cycle corresponding to $b_{[2,4]}$ appears in the lower-half plane $\text{Im}(z)<0$ in the figure (note that $b_{[2,4]}$ can move to the upper-half plane, then $b_{[1,3]}$ should appear in the lower-half plane).

\begin{figure}[h]
\centering
\includegraphics[width=9cm]{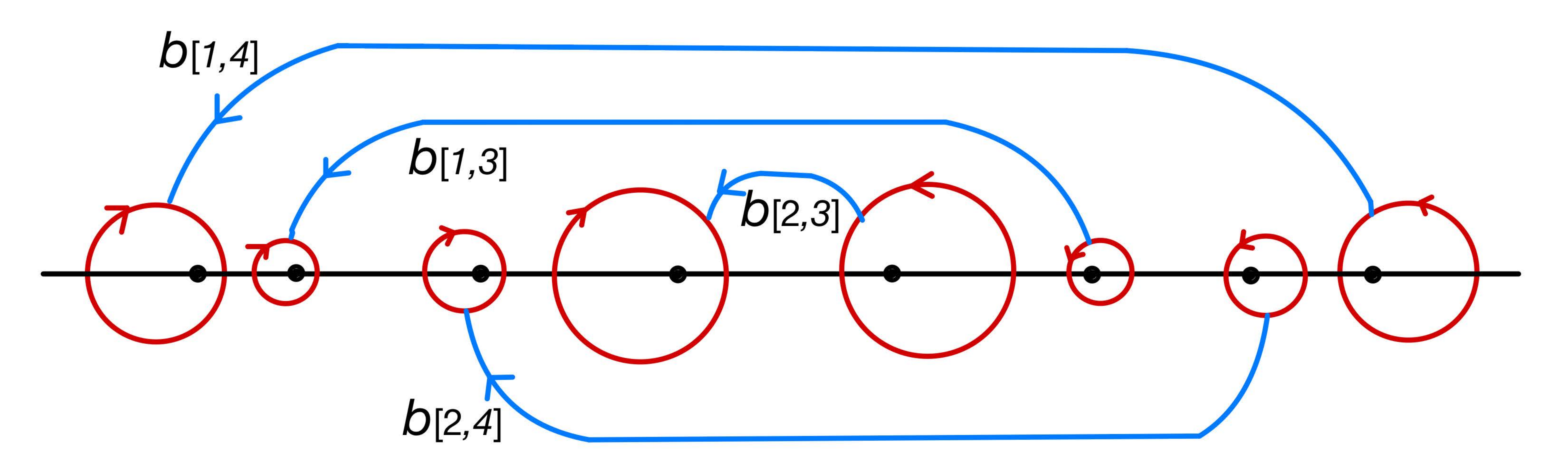}
\end{figure}

We consider the limit $\mu_p\to0$ but keep
all $\kappa_{k.l}$ distinct. Then from Theorem \ref{thm:theta-tau}, the corresponding $M$-theta function 
in \eqref{eq:Theta} is given by
\[
\widetilde{\theta}_4=1=\sum_{p=1}^4e^{\tilde{\phi}_p}+\sum_{p<q}C_{p,q}e^{\tilde{\phi}_p+\tilde{\phi}_q}
+\cdots +\left(\prod_{p<q}C_{p,q}\right)e^{\sum_{p=1}^4\tilde{\phi}_p},
\]
which consists of $2^4=16$ terms, and corresponds to 4-soliton solution of Hirota-type (see Remark \ref{rem:Hirota}).
One should note that the coefficient $C_{2,3}<0$, which implies the soliton solution given by the $M$-theta function above is \emph{not} regular (see Theorem \ref{thm:regularity}). Note also that the matrix $\widetilde{A}$ for this solution is given by
\[
\widetilde{A}=\begin{pmatrix}
1 &  & & & & & &  \tilde{A}_{1,4}\\
&1& & & & \tilde{A}_{1,3}& & \\
&&1&&& &\tilde{A}_{2,4}&\\
&&&1&\tilde{A}_{2,3}
\end{pmatrix}
\]
where $\tilde{A}_{i,j}$ are nonzero constants, and all other entries except pivots are zero. It is also easy to see that the matrix $\widetilde{A}$ is not TNN for any nonzero entries $\tilde{A}_{i,j}$ (see Remark \ref{rem:Hirota}, and also \cite{HKL:24}). This implies that the quasi-periodic solution associated with the Riemann surface uniformized by the Schottky group in the figure is not regular at least small parameters $\mu_p$.

By taking further limits $\alpha_p=\kappa_{i,j}\to\kappa_i$ and $\beta_p=\kappa_{j,i}\to\kappa_j$ with $p=\ell([i,j])$, we obtain the \emph{regular} solution with
\[
A=\begin{pmatrix}
1& 0& A_{1,3}&A_{1,4}\\
0& 1& A_{2,3}&A_{2,4}
\end{pmatrix}
\]
where $A_{1,3},A_{1,4}<0$, $A_{2,3},A_{2,4}>0$ and $A_{13}A_{2,4}-A_{2,3}A_{1,4}\ge0$, i.e. $A\in\text{Gr}(2,4)_{\ge0}$

\vskip0.3cm
\noindent
(b) {\bf A case with $g=3$:} Consider the $O\Le$-diagram \young[2,2][6][1, ,2,3], which gives
\[
P_1(A)=\{[1,4], \, [2,4], \, [2,3]\}.
\]
The Schottky parameters $\{\alpha_p,\beta_p; p=\ell([i,j])~\text{with}~[i,j]\in P_1(A)\}$ are given by
\[
\alpha_1=\kappa_{1,4}~<~\alpha_2=\kappa_{2,4}~<~\alpha_3=\kappa_{2,3}~<~\beta_3=\kappa_{3,2}~<~\beta_2=\kappa_{4,2}~<~\beta_1=\kappa_{4,1}.
\]

\begin{figure}[h]
\centering
\includegraphics[width=9cm]{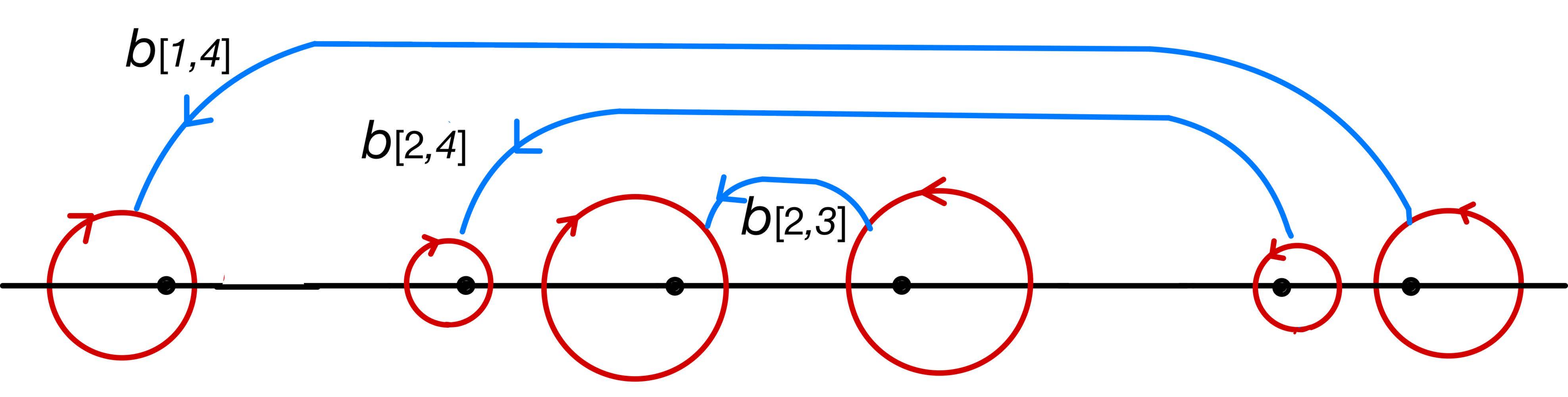}
\end{figure}

The limit with $\mu_p\to 0$ (keeping $\kappa_{i,j}$ distinct) gives the matrix
\[
\widetilde{A}=\begin{pmatrix}
1& & & & & & \tilde{A}_{1,4}\\
& 1& & & & \tilde{A}_{2,4}&\\
& & & 1 &\tilde{A}_{2,3}& & 
\end{pmatrix}
\]
which gives a 3-soliton solution without resonance (i.e. Hirota-type), and it is regular if
$\tilde{A}_{1,4}>0$, $\tilde{A}_{2,4}<0$ and $\tilde{A}_{2,3}>0$. After taking the limits $\kappa_{i,j}\to \kappa_i$,
the corresponding matrix $A\in\text{Gr}(2,4)_{\ge0}$ is 
\[
A=\begin{pmatrix}
1&0&0&A_{1,4}\\
0&1&A_{2,3}&A_{2,4}
\end{pmatrix}
\]
where $A_{1,4}<0$ and $A_{2,3},A_{2,4}>0$. We also note that the quasi-periodic solution is regular, and
the Riemann surface in this case is an $\mathsf{M}$-curve of genus $3$. Notice that the $b$-cycles 
corresponding to $b_{[i,j]}$'s are all in the upper-half plane $\text{Im}(z)\ge0$.

In the similar argument, we can construct other type KP soliton from $\text{Gr}(2,4)_{\ge0}$, for example,
consider the case with the $O\Le$-diagram \young[2,2][6][1,2,,3] or \young[2,1][6][1,2,3]. The Riemann surfaces for these examples are $\mathsf{M}$-curves.

\subsection{A case in $\text{Gr}(5,9)_{\ge0}$}
Here we consider Example \ref{ex:P1}. The Schottky parameters $\{\alpha_p,\beta_p\}$ for $1\le p\le 10$ are given in the table:
\begin{table}[h]
			\centering
			\begin{tabular}{|c||c|c|c|c|c|c|c|c|c|c|}
				\hline
				
				$p$ & $1$ & $2$ &$3$ &$4$ & $5$ & 6 & 7 & 8 & 9 & 10 \\  
				\hline 
				$\alpha_p$   & $\kappa_{1,9}$ & $\kappa_{1,8}$ & $\kappa_{2,9}$ & $\kappa_{2,8}$&$\kappa_{2,3}$  & $\kappa_{4,5}$ &$\kappa_{6,9}$ &$\kappa_{6,8}$ &$\kappa_{7,9}$ & $\kappa_{7,8}$  \\ 
				\hline  
				$\beta_p$ & $\kappa_{9,1}$ & $\kappa_{8,1}$  & $\kappa_{9,2}$ & $\kappa_{8,2}$ & $\kappa_{3,2}$
	& $\kappa_{5,4}$ &$\kappa_{9,6}$& $\kappa_{8,6}$ &$\kappa_{9,7}$ & $\kappa_{8,7}$ \\   
				\hline
			\end{tabular}
		\end{table}
		
\noindent	
The fundamental domain $\mathcal{F}(\Gamma_a)$ and the homological cycles are shown in the figure below.
\begin{figure}[h]
\centering
\includegraphics[width=11cm]{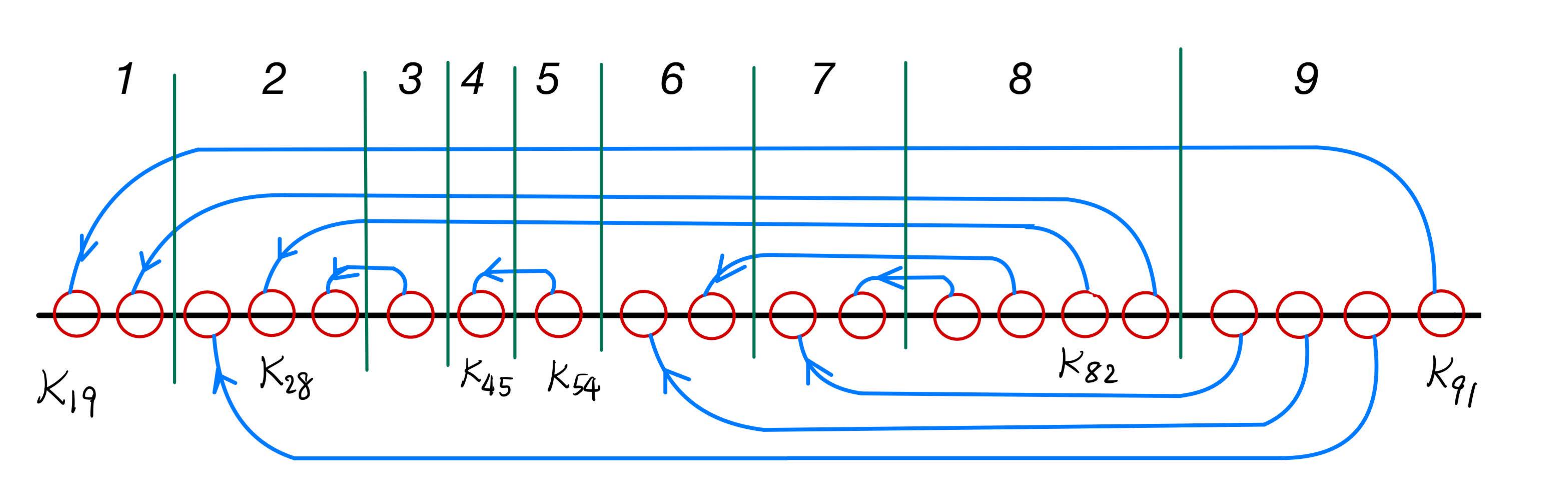}
\end{figure}

In the limits $\mu_p\to 0$ keeping all $\kappa_{i,j}$ distinct, we have 10 line solitons of Hirota-type,
whose $M$-theta function in Theorem \ref{thm:theta-tau} has negative coefficients, $C_{2,3}, ~C_{2,7}, ~C_{2,9}, ~C_{4,7}, ~C_{4,9}, ~C_{8,9}~<~0$ as shown in Example \ref{ex:C59}. This implies that this Hirota-type 10-soliton solution is 
singular, that is, the corresponding matrix $A$ is \emph{not} TNN. And the quasi-periodic solution associated with the Riemann surface uniformized by the Schottky group is not regular at least for small $\mu_p$.

However, in the further limits $\alpha_p=\kappa_{i,j}\to\kappa_i, \beta_p=\kappa_{j,i}\to\kappa_j$ with $p=\ell([i,j])$, the quasi-periodic solution \eqref{eq:QPS} converges to the KP soliton in Theorem \ref{thm:theta-tau}. Here note that from the $O\Le$-diagram and Proposition \ref{prop:Cpq}, we have, for example,
\[
\phi_1+\phi_4=(\xi_9-\xi_1)+(\xi_8-\xi_2)=(\xi_8-\xi_1)+(\xi_9-\xi_2)=\phi_2+\phi_3,
\]
where $\xi_j=\kappa_j x+\kappa_j^2y+\kappa_j^3t$. For the regular soliton, one need to take 
$C_{1,4}e^{\phi_1^0+\phi_4^0}+C_{2,3}e^{\phi_2^0+\phi_3^0}\ge 0$, $C_{1,8}e^{\phi_1^0+\phi_8^0}+C_{2,7}e^{\phi_2^0+\phi_7^0}\ge 0$ etc (see Proposition \ref{prop:Cpq} and Subsection \ref{sec:Ex24}). That is, the negative coefficients $C_{p,q}$ are cancelled by appropriate choices of the parameters $\phi_p^0$'s.

It is interesting note that if the boxes $3,7$ and $9$ in the $O\Le$-diagram are all empty, then
we get a regular soliton solution as well as regular quasi-periodic solution with genus 7. This corresponds to 
add zero at the $(7,9)$ box in the $\Le$-diagram. In this case, all the $b$-cycles appear in the upper-half plane $\text{Im}(z)\ge 0$. It can be seen that the regular quasi-periodic solution is obtained when 
all the $b$-cycles are in the upper half plane (this corresponds to a non-crossing permutation
for the regularity \cite{HKL:24}).


\bigskip
\noindent
{\bf Acknowledgements.}
One of the authors (TI) is partially supported by the JSPS Grant-in-Aid for Scientific Research No. 25K06920.

 \bigskip
 \noindent
 {\bf Declarations.}
 \begin{itemize}
\item Data sharing not applicable to this article as no datasets were generated or analyzed during the current study.\\
 \item The author declares no conflicts of interest associated with this manuscript.
 \end{itemize}


\end{document}